%% file: thesis.tex
\documentclass[12pt]{report}   
\usepackage{graphicx}  
\usepackage[letterpaper, left=1in, right=1in, top=1in, bottom=1in]{geometry}
\usepackage{enumitem}
\usepackage{xltabular}
\usepackage{booktabs,tabularx}
\usepackage{amssymb}
\usepackage{setspace}  
\usepackage{times}  
\usepackage[explicit]{titlesec}  
\usepackage[titles]{tocloft}  
\usepackage{siunitx}  
\usepackage{xspace}  
\usepackage[backend=bibtex, sorting=none, bibstyle=ieee]{biblatex}  
\usepackage{appendix}  
\usepackage{rotating}  
\usepackage[normalem]{ulem}  
\usepackage{textcomp} 
\usepackage{indentfirst} 
\usepackage{booktabs,array,arydshln} 
\usepackage{amsmath} 
\usepackage[T1]{fontenc} 
\usepackage[utf8]{inputenc} 
\usepackage{newtxmath} 
\usepackage[hidelinks]{hyperref}
\usepackage{graphicx}
\usepackage{float}
\usepackage{tikz}
\usepackage{amsmath}
\usepackage{graphicx}
\usetikzlibrary{arrows.meta, positioning, shapes, calc}

\usepackage[ruled,vlined,linesnumbered]{algorithm2e}
\DontPrintSemicolon
\SetKwInOut{Input}{Input}
\SetKwInOut{Output}{Output}
\SetKwProg{Fn}{Function}{ is}{end}
\SetKw{KwRet}{return}
\SetKwComment{Comment}{$\triangleright$ }{}
\SetAlgoInsideSkip{smallskip}
\SetAlgoCaptionSeparator{:}
\RestyleAlgo{ruled}
\bibliography{references}

\AtEveryBibitem{\clearfield{issn}}
\AtEveryBibitem{\clearlist{issn}}

\AtEveryBibitem{\clearfield{language}}
\AtEveryBibitem{\clearlist{language}}

\AtEveryBibitem{\clearfield{doi}}
\AtEveryBibitem{\clearlist{doi}}

\AtEveryBibitem{\clearfield{url}}
\AtEveryBibitem{\clearlist{url}}

\AtEveryBibitem{%
  \ifentrytype{online}
    {}
    {\clearfield{urlyear}\clearfield{urlmonth}\clearfield{urlday}}}

\begin{document}
\doublespacing  

\newcommand{\evoname}{Evo-1-8k-base\xspace}
\newcommand{\ampN}{\num{3200}}
\newcommand{\ampSpecies}{\num{27}}


\phantomsection
\addcontentsline{toc}{chapter}{Preface}
\input{Preface.tex}

\pagenumbering{gobble}
\input{Abstract.tex}

\pagenumbering{roman}
\setcounter{page}{1} 
\renewcommand{\cftchapdotsep}{\cftdotsep}  
\renewcommand{\cftchapfont}{\normalfont}  
\renewcommand{\cftchappagefont}{}  
\renewcommand{\cftchappresnum}{Chapter }
\renewcommand{\cftchapaftersnum}{:}
\renewcommand{\cftchapnumwidth}{5em}
\renewcommand{\cftchapafterpnum}{\vskip\baselineskip} 
\renewcommand{\cftsecafterpnum}{\vskip\baselineskip}  
\renewcommand{\cftsubsecafterpnum}{\vskip\baselineskip} 
\renewcommand{\cftsubsubsecafterpnum}{\vskip\baselineskip} 

\titleformat{\chapter}[display]
{\normalfont\bfseries\filcenter}{\chaptertitlename\ \thechapter}{0pt}{\large{#1}}

\renewcommand\contentsname{Table of Contents}

\begin{singlespace}
\tableofcontents
\setlength{\cftparskip}{\baselineskip}
\listoffigures
\listoftables
\end{singlespace}
\clearpage
\input{abbreviations.tex}

\clearpage

\phantomsection
\addcontentsline{toc}{chapter}{Acknowledgments}
\input{acknowledgements.tex}




\clearpage
\pagenumbering{arabic}
\setcounter{page}{1}

\titleformat{\chapter}[display]
{\normalfont\bfseries\filcenter}{}{0pt}{\large\chaptertitlename\ \large\thechapter : \large\bfseries\filcenter{#1}}  
\titlespacing*{\chapter}
  {0pt}{0pt}{30pt}	
  
\titleformat{\section}{\normalfont\bfseries}{\thesection}{1em}{#1}

\titleformat{\subsection}{\normalfont}{\thesubsection}{0em}{\hspace{1em}#1}



\input{chapter1.tex}




\input{chapter3.tex}

 
\input{chapter4.tex}
\input{chapter5.tex}
\input{chapter6.tex}

\clearpage
\phantomsection 
\input{Conclusion.tex}

\clearpage
\phantomsection 
\titleformat{\chapter}[display]
{\normalfont\bfseries\filcenter}{}{0pt}{\large\bfseries\filcenter{#1}}  
\titlespacing*{\chapter}
  {0pt}{0pt}{30pt}

\begin{singlespace}  
	\setlength\bibitemsep{\baselineskip}  
	\addcontentsline{toc}{chapter}{References}  
	\printbibliography[title={References}]
\end{singlespace}


\titleformat{\chapter}[display]
{\normalfont\bfseries\filcenter}{}{0pt}{\large\chaptertitlename\ \large\thechapter : \large\bfseries\filcenter{#1}}  
\titlespacing*{\chapter}
  {0pt}{0pt}{30pt}	
  
\titleformat{\section}{\normalfont\bfseries}{\thesection}{1em}{#1}

\titleformat{\subsection}{\normalfont}{\thesubsection}{0em}{\hspace{1em}#1}

\appendix
\input{appendix_phylo.tex}

\end{document}

%% file: Preface.tex
\begin{center}
\vspace*{2\baselineskip}

{\LARGE Cross-Species Antimicrobial Resistance Prediction from Genomic Foundation Models\\ }
 
\vspace{2\baselineskip}

Huilin Tai

\vspace{0.5\baselineskip}

uni: ht2666  

\vspace{1\baselineskip}

Advisor: Dr. Mohammed AlQuraishi

Committee:  Dr. David Knowles, Dr. Tal Korem   

\vspace{2\baselineskip}

Submitted in partial fulfillment of the\\
requirements for the degree\\
of Master of Science\\
in the Department of Computer Science at the School of Engineering and Applied Science

\vspace{1\baselineskip}

COLUMBIA UNIVERSITY

\vspace{0.5\baselineskip}

2025  

\end{center}

%% file: Abstract.tex
\begin{abstract}
Cross-species antimicrobial resistance (AMR) prediction is fundamentally an out-of-distribution generalization problem: models trained on one set of bacterial taxa must transfer to phylogenetically distinct genomes that may rely on different resistance mechanisms. Critically, resistance is not monolithic. Across species, it arises from a heterogeneous mixture of localized, horizontally transferred gene cassettes and diffuse, species-specific genomic backgrounds, making successful transfer inherently mechanism-dependent.

Using a strict \emph{species holdout} protocol, we first establish an interpretable k-mer baseline with \textit{Kover}, showing that strong within-species performance collapses under true cross-species evaluation. This motivates the need for representation-level choices that explicitly preserve transferable biological signals rather than amplify phylogenetic shortcuts.

We introduce two ingredients that make genomic foundation model embeddings effective for cross-species AMR prediction. First, for layer selection, we develop diagnostics for activation scale, isotropy, effective rank, and cross-seed stability under native bfloat16 inference. These reveal a sharp stability boundary at Layer~11 in \evoname, identifying Layer~10 as the deepest jointly stable layer; extracting embeddings here improves downstream conditioning, reproducibility, and robustness.

Second, for feature aggregation, we argue that global pooling obscures localized resistance mechanisms. Instead, we treat per-window embeddings as an ordered multivariate signal and apply MiniRocket to summarize multi-scale local activation patterns. This preserves cassette-scale signals (e.g., plasmid-borne $\beta$-lactamases) that global averages dilute, reorganizing feature space toward phenotype-aligned neighborhoods where simple classifiers can generalize across species.

On ampicillin resistance across 3{,}388 genomes from 126 species, we show that cross-species performance depends on which resistance mechanisms dominate the held-out species, not on aggregation method alone. MiniRocket excels when cassette-mediated resistance predominates; Global Pooling remains competitive for chromosomal or diffuse mechanisms. Both approaches perform similarly under same-species evaluation.

Beyond accuracy, MiniRocket enables zero-training aggregation, interpretable predictions via neighbor auditing, and biological validation through mechanism-based clustering. Unlike complex decision boundaries learned by gradient boosting, k-NN exposes the underlying geometric reorganization that explains when and why local pattern preservation succeeds: reduced phylogenetic hubness and increased cross-species mechanism sharing. Together, our results establish aggregation choice as a central axis in cross-species AMR prediction and provide a reproducible, diagnostic-driven framework for deploying genomic foundation models under distribution shift.
\end{abstract}

%% file: abbreviations.tex
\phantomsection
\addcontentsline{toc}{section}{List of Abbreviations}
\section*{List of Abbreviations}
\begin{flushleft}
\small
\begin{xltabular}{\textwidth}{@{} >{\raggedright\arraybackslash}p{0.18\textwidth} >{\raggedright\arraybackslash}X @{}}
\toprule
\textbf{Abbreviation} & \textbf{Definition} \\
\midrule
\endfirsthead

\multicolumn{2}{c}{\small\textit{List of Abbreviations (continued)}} \\[0.5ex]
\toprule
\textbf{Abbreviation} & \textbf{Definition} \\
\midrule
\endhead

\bottomrule
\endfoot

AMR & Antimicrobial Resistance \\
AST & Antimicrobial Susceptibility Testing \\
AUROC & Area Under the Receiver Operating Characteristic Curve \\
AUPRC & Area Under the Precision--Recall Curve \\
bf16 & bfloat16 (Brain Floating Point 16-bit) \\
BV-BRC & Bacterial and Viral Bioinformatics Resource Center \\
CARD & Comprehensive Antibiotic Resistance Database \\
CLIA & Clinical Laboratory Improvement Amendments \\
DNA & Deoxyribonucleic Acid \\
ECE & Expected Calibration Error \\
Evo & Evo-1-8k-base (genomic foundation model) \\
GC & Guanine--Cytosine \\
GTDB-Tk & Genome Taxonomy Database Toolkit \\
HMM & Hidden Markov Model \\
IoU & Intersection over Union \\
IRB & Institutional Review Board \\
k-NN & k-Nearest Neighbors \\
L10, L11 & Layer 10, Layer 11 (neural network layers) \\
LLM & Large Language Model \\
LOSO & Leave-One-Species-Out \\
MCC & Matthews Correlation Coefficient \\
MIC & Minimum Inhibitory Concentration \\
MiniRocket & MiniRocket (time-series classification method) \\
NLP & Natural Language Processing \\
OOD & Out-of-Distribution \\
PATRIC & PAThosystems Resource Integration Center \\
PCA & Principal Component Analysis \\
PPV & Proportion of Positive Values \\
RNA & Ribonucleic Acid \\
ROCKET & Random Convolutional Kernel Transform \\
SCM & Set Covering Machine \\
SNP & Single Nucleotide Polymorphism \\
SRP & Sparse Random Projection \\
SVM & Support Vector Machine \\
ulp & Unit in the Last Place (floating-point precision) \\
WGS & Whole Genome Sequencing \\
\end{xltabular}
\end{flushleft}

%% file: acknowledgements.tex
\section*{Acknowledgements}
I would like to thank my advisor, Prof.\ Mohammed AlQuraishi, for his guidance, feedback, and support throughout this project. I am especially grateful to Harry Lee, Ph.D.\ student in the AlQuraishi Lab, for meeting with me weekly, reviewing my ideas and code, brainstorming directions, and providing implementation support when I was blocked. I also thank the researchers in the Genomic Language Modeling (GLM) group at the AlQuraishi Lab; I learned a great deal from their presentations and feedback in our weekly meetings. I joined the lab at a personal transition point, unsure of my research direction, and I am grateful for the space and platform to learn, explore, and find my path. Finally, I thank the AlQuraishi Laboratory at Columbia University for the computational resources that made these experiments possible.

%% file: chapter1.tex
\chapter{Introduction}
\label{chap:intro}

Antimicrobial resistance (AMR) causes over 1.27 million deaths annually \cite{Murray2022} and is projected to reach up to 10 million deaths per year by 2050 \cite{WHO2019,ONeill2016}. Current culture-based susceptibility testing requires 48 to 72 hours \cite{CLSI2020}, forcing clinicians either to delay treatment or to prescribe broad-spectrum antibiotics empirically. Both options carry substantial risk. Although modern sequencing technologies can decode bacterial genomes within hours, translating genomic sequences into accurate and reliable resistance predictions remains computationally and biologically challenging.

This thesis addresses two central questions. 
First, how can whole-genome embeddings produced by genomic foundation models 
be used effectively under realistic distribution shift? 
Second, what biological structure is reflected in the learned feature spaces 
of these models, and how does this structure constrain cross-species 
generalization?

The computational challenge is immediate. Genomic foundation models produce embeddings of extremely high dimensionality when applied to bacterial genomes. Consider Evo-1-8k-base \cite{Nguyen2024}, the model used throughout this thesis. A typical four-megabase bacterial genome requires approximately one thousand non-overlapping four-kilobase windows for full coverage. Each window produces a 4,096-dimensional embedding, yielding more than four million raw features per genome. This scale far exceeds the capacity of standard machine learning pipelines and makes naive downstream modeling impractical.

Beyond dimensionality, the biological structure of antimicrobial resistance introduces a deeper difficulty. Resistance determinants are sparse and heterogeneous. For example, a three-kilobase $\beta$-lactamase cassette occupies less than 0.1 percent of a typical bacterial genome, while other resistance phenotypes arise from diffuse chromosomal mutations that alter regulation, membrane permeability, or target affinity. Aggregation strategies adapted from natural language processing assume relatively uniform information density and therefore tend to dilute sparse but functionally critical genomic signals. This thesis develops methods that preserve biologically meaningful local structure in genomic embeddings, enabling cross-species resistance prediction without sacrificing interpretability.

\section{The Challenge of Cross-Species Prediction}

Cross-species AMR prediction is fundamentally an out-of-distribution generalization problem. Training on genomes from one set of bacterial species and evaluating on phylogenetically distinct species induces severe covariate shift. Bacterial genomes differ substantially in GC content, codon usage bias, and chromosomal organization \cite{Nishida2012}. These compositional properties correlate strongly with species identity and often dominate learned representations.

As a result, models trained on genomic data inevitably entangle resistance mechanisms with species-specific background signals. During training, this entanglement is advantageous. GC content and k-mer frequencies provide strong predictive cues and improve apparent accuracy. Under species shift, however, these cues no longer transfer. A model trained primarily on Enterobacterales implicitly learns their characteristic genomic signatures. When evaluated on Firmicutes, these species-specific patterns become uninformative even if identical resistance mechanisms are present.

Complicating matters further, identical resistance phenotypes can arise from distinct genomic mechanisms. For example, \emph{Escherichia coli} typically achieves ampicillin resistance through plasmid-encoded $\beta$-lactamases, whereas \emph{Pseudomonas aeruginosa} relies on chromosomal \textit{ampC} induction combined with membrane modifications. The challenge is therefore not whether models can learn resistance, but whether they can isolate transferable functional elements from species-specific genomic context.

\paragraph{Mechanism-mix hypothesis.}
The central hypothesis of this thesis is that resistance mechanisms differ in their transferability across species. Horizontally transferred resistance cassettes are subject to strong purifying selection and therefore maintain sequence conservation across taxonomic boundaries. In contrast, chromosomal resistance mechanisms are embedded in species-specific regulatory and structural contexts. This predicts that aggregation strategies preserving localized genomic signals should benefit cassette-mediated resistance in particular, rather than improving cross-species prediction uniformly.

\section{Problem Formalization}

We formalize the cross-species AMR prediction task as follows. Given a collection of genome sequences $\mathcal{X} = \{x_i\}_{i=1}^N$, where each genome $x_i \in \Sigma^{L_i}$ is defined over the DNA alphabet $\Sigma = \{A, C, G, T\}$ with length $L_i$, and corresponding binary resistance phenotypes $\mathcal{Y} = \{y_i\}_{i=1}^N$ for a given antibiotic, the goal is to learn a function $f : \mathcal{X} \rightarrow [0,1]$ that predicts the probability of resistance.

The critical evaluation constraint is species holdout partitioning. Let $s_i$ denote the species label of genome $i$. We require that the species sets of the training and test splits satisfy $S_{\text{train}} \cap S_{\text{test}} = \emptyset$. This induces covariate shift \cite{Shimodaira2000} such that $P_{\text{train}}(x) \neq P_{\text{test}}(x)$ due to phylogenetic divergence. Standard supervised learning methods implicitly assume stable input distributions. When this assumption fails, models often rely on spurious correlations that do not generalize.

Our working hypothesis is that certain functional elements, particularly horizontally transferred resistance cassettes, retain sufficient sequence conservation to remain identifiable across species boundaries. $\beta$-lactamase genes exemplify this phenomenon. Mutations in catalytic residues abolish enzymatic function, creating strong selective pressure for sequence preservation. These genes therefore exhibit higher conservation than their surrounding genomic context and may be recoverable under species shift if properly isolated.

Most prior work evaluates AMR prediction using random or stratified train-test splits that permit phylogenetic overlap between training and test sets. This allows models to exploit within-species similarity rather than learning truly transferable resistance mechanisms \cite{Hu2024}. Species holdout evaluation eliminates this shortcut and provides a more realistic assessment of cross-species generalization.

\section{Establishing the Baseline: Kover and Out-of-Distribution Testing}

This investigation begins by establishing rigorous baselines using Kover \cite{Drouin2016}, an interpretable method that learns sparse Boolean rules over k-mer presence and absence. Kover has been reported to achieve strong within-species accuracy, typically between 85 and 92 percent \cite{Drouin2019}, making it a suitable baseline for comparison.

To evaluate generalization under distribution shift, we designed an out-of-distribution testing framework based on strict species holdout partitioning. This evaluation strategy is largely absent from the AMR prediction literature. Standard practice relies on random train-test splits or stratified sampling that preserves species proportions. Such protocols do not reflect the clinical objective, which is to generalize to previously unseen pathogens rather than to new isolates of familiar species. Species holdout evaluation enforces zero phylogenetic overlap between training and test sets and more accurately reflects real-world deployment.

Applying this framework to Kover reveals substantial and inconsistent performance degradation. Within-species F1 scores remained stable (0.68–0.84), but cross-species F1 ranged from 0.02 to 0.87 depending on which species were held out, with performance strongly influenced by the dominant resistance mechanisms in each held-out set. Examination of learned rules shows that Kover relies heavily on species-specific k-mer s. A 31-mer predictive of resistance in \emph{E. coli} may be entirely absent in \emph{Klebsiella}, even when both species share the same $\beta$-lactamase genes. Kover captures local sequence context rather than conserved functional elements. This pattern is consistent across six antibiotics, indicating that strong within-species performance does not imply robustness under species shift.

\section{Genomic Foundation Models and the Scale Problem}

Genomic foundation models such as Evo \cite{Nguyen2024} offer a potential alternative. By combining attention mechanisms with state-space modeling, Evo achieves near-linear scaling and supports context windows of up to 131 kilobases, sufficient to capture complete resistance operons. Pretraining on diverse genomic data suggests that Evo may encode general sequence-function relationships that transfer across species.

A standard embedding-based pipeline extracts Evo embeddings from a fixed layer, reduces dimensionality using principal component analysis retaining 90 percent of variance, and aggregates window-level embeddings using summary statistics such as mean, standard deviation, minimum, maximum, and interquartile values. However, the scale problem remains fundamental. A four-megabase genome still produces on the order of one thousand window embeddings of 4,096 dimensions each. Practical deployment therefore hinges on two design choices: which layer to extract from, and how to aggregate window embeddings without destroying sparse local signals.

\section{Two Key Innovations}

\paragraph{Layer-wise diagnostics.}
Rather than defaulting to the final model layer, we evaluate all 32 Evo layers using diagnostics that include activation magnitude, isotropy, effective rank, self-similarity, and downstream task stability. This analysis reveals numerical and representational degradation beyond Layer 11 and identifies Layer 10 as the deepest jointly stable extraction point. All downstream analyses in this thesis use embeddings extracted from Layer 10.

\paragraph{Local-pattern-preserving aggregation.}
To preserve localized resistance signals, we treat the ordered sequence of window embeddings as a multivariate signal and apply MiniRocket \cite{Dempster2021}. MiniRocket computes large banks of simple order-sensitive features by applying random convolutions followed by proportion-of-positive-values pooling. Unlike global statistical aggregation, this approach preserves sparse cassette-scale patterns while down-weighting diffuse species-specific context.

Each kernel computes the proportion of positive values as
\[
\text{PPV}_k = \frac{1}{T - \ell_k + 1} \sum_{t=1}^{T - \ell_k + 1}
\mathbb{I}\!\left[\sum_{j=0}^{\ell_k-1} \mathbf{w}_k^\top \mathbf{h}_{t+j} > 0\right].
\]

\section{Results and Mechanistic Understanding}

Combining Layer 10 extraction with MiniRocket aggregation reveals that cross-species AMR performance is mechanism-dependent. On ampicillin resistance across 3{,}200 genomes from 126 species under strict species holdout evaluation, MiniRocket performs best when cassette-mediated resistance predominates, whereas Global Pooling remains competitive for chromosomal or diffuse resistance mechanisms. Both approaches perform similarly under same-species evaluation.

Quantitatively, both MiniRocket and Global Pooling substantially outperform 
the k-mer  baseline (Kover) under strict species holdout, but their gains 
reflect different inductive biases rather than uniform superiority. Without 
any explicit stratification by resistance mechanism, MiniRocket-based models 
achieve high F1 scores on cross-species validation (\texttt{val\_outside}), 
with k-NN reaching F1 = 0.982 compared to F1 = 0.901 for Global Pooling k-NN 
and severe degradation for Kover (F1 dropping from approximately 0.68 within 
species to 0.02 under species holdout). On the independent cross-species test 
set (\texttt{test\_outside}), which happens to include a larger fraction of 
species dominated by chromosomal or diffuse resistance mechanisms, Global 
Pooling combined with linear or tree-based classifiers achieves comparable 
or higher F1 scores than MiniRocket.

Several observations support this interpretation. Species dominated by plasmid-mediated resistance show higher cross-species accuracy than those relying on chromosomal mechanisms. Post-transformation embedding geometry clusters more strongly by resistance gene family than by phylogeny. Antibiotics whose resistance is primarily cassette-mediated exhibit stronger cross-species transfer than those dominated by point mutations.

These results also clarify limitations. Chromosomal resistance remains difficult to predict under species shift. Performance depends on phylogenetic diversity in the training set and does not extrapolate to entirely novel resistance mechanisms. All analyses are retrospective, and prospective clinical validation is required before deployment.

\section{Contributions}

This thesis makes four primary contributions:

\begin{enumerate}
  \item A species holdout evaluation protocol and associated infrastructure for leakage-resistant benchmarking of AMR prediction models.
  \item A diagnostic framework for layer selection in genomic foundation models, identifying Layer 10 as a stable and transferable extraction point.
  \item A local-pattern-preserving aggregation strategy that reveals when and why local signals enable cross-species AMR prediction, outperforming Global Pooling for cassette-mediated mechanisms while remaining comparable for chromosomal resistance.
  \item Empirical evidence that cross-species AMR prediction can be competitive when resistance mechanisms are modular and embeddings are extracted and aggregated appropriately.
\end{enumerate}

\section{Thesis Organization}

Chapter~\ref{chap:data} describes dataset construction, quality control, and species holdout partitioning. Chapter~\ref{chap:kover} presents systematic benchmarking that exposes catastrophic cross-species failure in traditional methods. Chapter~\ref{chap:evo_embedding} develops the layer-wise diagnostic framework. Chapter~\ref{chap:amr_prediction} reports the main experimental results, including geometric analyses.

This thesis demonstrates that effective use of genomic foundation models requires understanding both their computational properties and the biological structure of antimicrobial resistance. By addressing numerical stability, preserving local genomic signals, and respecting evolutionary constraints, we achieve cross-species resistance prediction that is both accurate and interpretable.

%% file: chapter3.tex
\chapter{Data}
\label{chap:data}

This chapter details the dataset and protocol. Instead of \emph{ad hoc} inputs, we build a reproducible, auditable corpus to test a single question: do foundation-model genome embeddings transfer to unseen categories? The design emphasizes traceability, balanced partitions, and intentional distribution shifts.

\section{Sources and Provenance}

We draw primary metadata from the BV–BRC command-line interface, including genome-level assembly statistics (e.g., N50, contig counts), CheckM completeness and contamination, and hierarchical taxonomies, together with binary antimicrobial susceptibility calls for genome–antibiotic pairs spanning 47 drugs. Species identifiers are harmonized against Entrez-derived mappings, and phylogeny is standardized with GTDB-Tk (v2.1.1). Sequences are retrieved via the BV–BRC FTP API: nucleotide (\texttt{.fna.gz}) and protein (\texttt{.faa.gz}) files for \num{89451} genomes are downloaded with exponential-backoff retries and MD5 verification. Median genome sizes are approximately 4.2~Mb (nucleotide) and 1.8~Mb (protein).  

\section{Initial Exploration and Motivation for Filtering}

\begin{figure}[ht]
  \centering
  \includegraphics[width=0.9\textwidth]{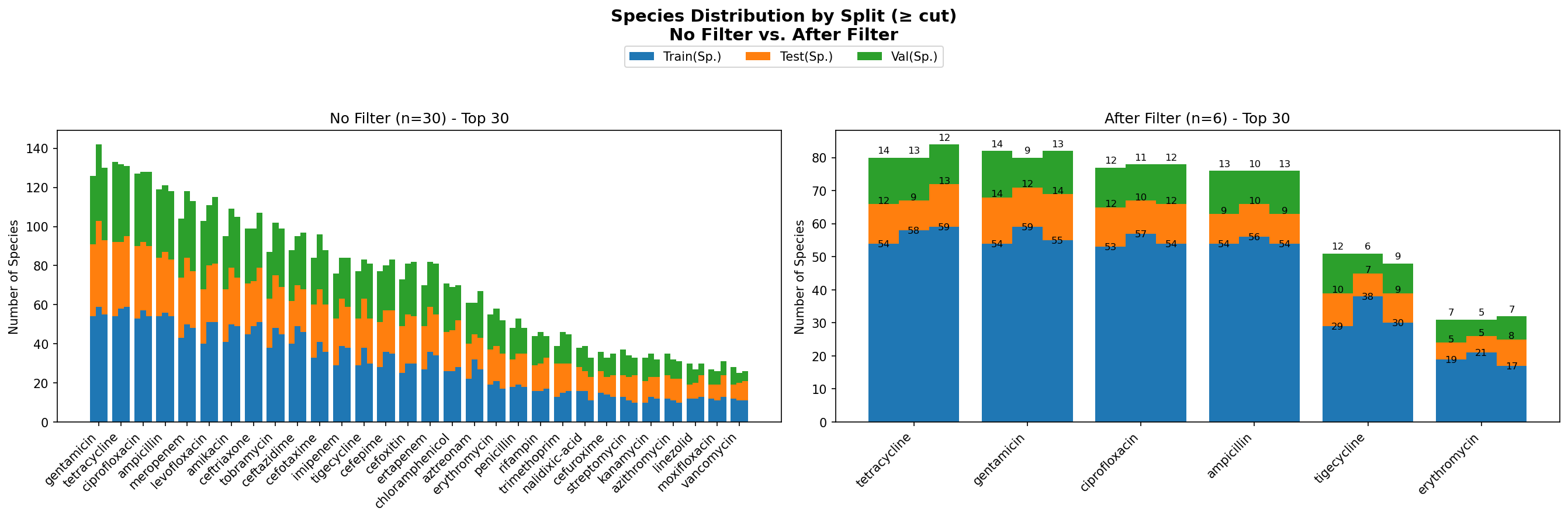}
  \caption{Category coverage among the top 30 targets, before (left) and after (right) filtering. The raw data shows extreme imbalance, with many antibiotics lacking sufficient species diversity. Post-filtering retains only targets that preserve diversity across all five partitions.}
  \label{fig:species_split_comparison}
\end{figure}

\begin{figure}[ht]
  \centering
  \includegraphics[width=0.9\textwidth]{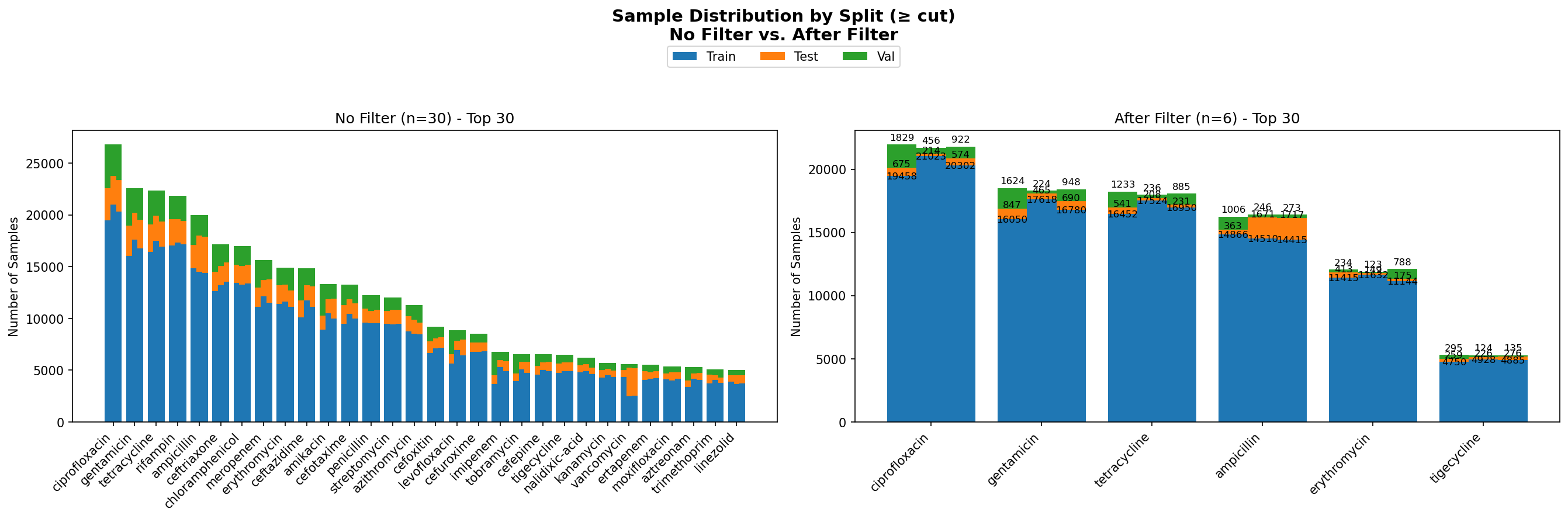}
  \caption{Sample counts among the top 30 targets, before (left) and after (right) filtering. Many antibiotics in the raw data have insufficient samples for stable evaluation. Post-filter sets maintain adequate per-partition counts to yield reliable confidence intervals.}
  \label{fig:samples_split_comparison}
\end{figure}

Before applying retention criteria, we examine the raw data landscape across 47 antimicrobial targets. Figures~\ref{fig:species_split_comparison} and~\ref{fig:samples_split_comparison} reveal substantial heterogeneity in species coverage and sample counts across different antibiotics. Many targets exhibit extreme skew—either dominated by a single species or containing insufficient positive examples for reliable evaluation. This exploration motivates strict filtering to ensure statistical power and category diversity in all evaluation partitions.

\section{Retention Criteria}

\subsection{Filtering Thresholds}

Based on the exploratory analysis, we establish strict retention criteria to ensure both statistical reliability and taxonomic diversity. An antibiotic was retained only if it met \emph{all} of the following thresholds:
\begin{itemize}
  \item At least 100 resistant samples in both validation and test sets
  \item At least 5 distinct species represented in both validation and test sets
  \item A minimum of 15,000 total samples across the entire dataset
\end{itemize}

These thresholds ensure statistical power for reliable AUROC/AUPRC estimation while preventing dominance by ubiquitous pathogens such as \emph{Escherichia coli} or \emph{Klebsiella pneumoniae}. The species diversity requirement guarantees that models cannot achieve high performance by memorizing patterns from a single dominant organism.

\subsection{Final Antibiotics Selected}

Applying these criteria to the 47 candidate antibiotics, exactly six satisfied all requirements:
\begin{description}
  \item[Ciprofloxacin:] A fluoroquinolone with broad activity against Gram-negative bacteria
  \item[Gentamicin:] An aminoglycoside targeting protein synthesis
  \item[Tetracycline:] A broad-spectrum bacteriostatic agent
  \item[Ampicillin:] A $\beta$-lactam with high resistance prevalence
  \item[Erythromycin:] A macrolide primarily active against Gram-positive bacteria
  \item[Tigecycline:] A glycylcycline with low resistance rates
\end{description}

Each demonstrated rich phenotype diversity and high representation across bacterial taxa, enabling robust assessment of both within-species and cross-species generalization.

\section{Label Distribution in the Filtered Corpus}

\begin{figure}[ht]
  \centering
  \includegraphics[width=0.9\textwidth]{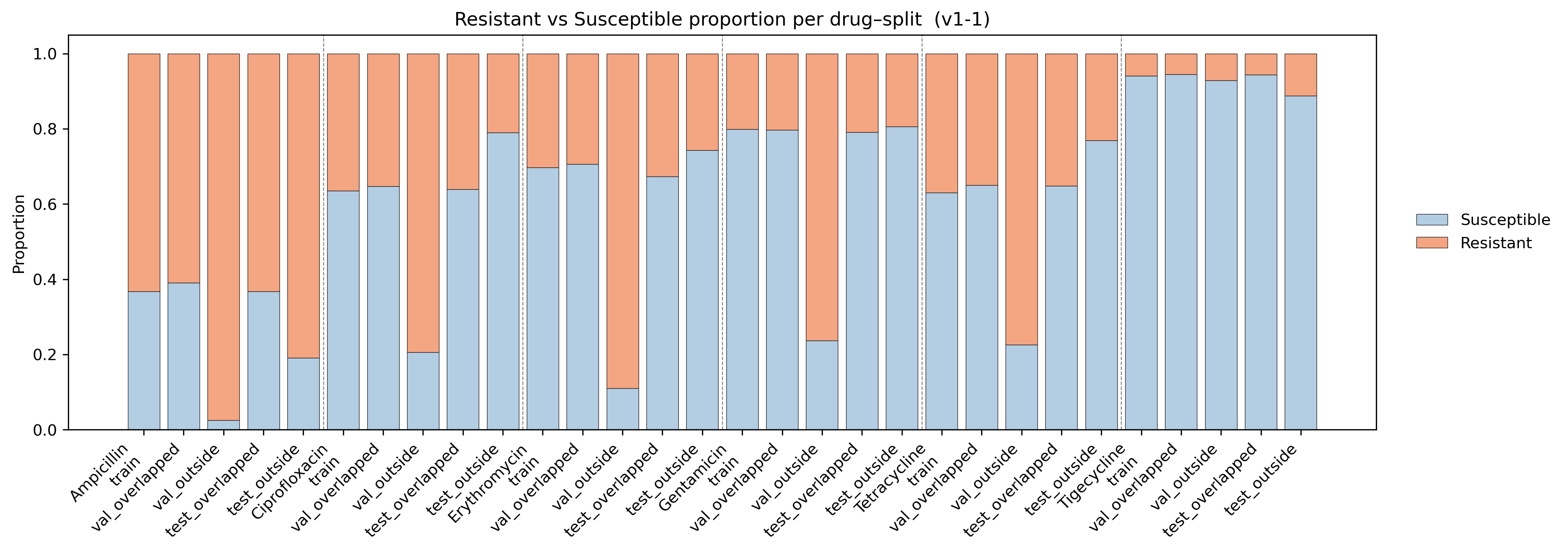}\\[6pt]
  \includegraphics[width=0.9\textwidth]{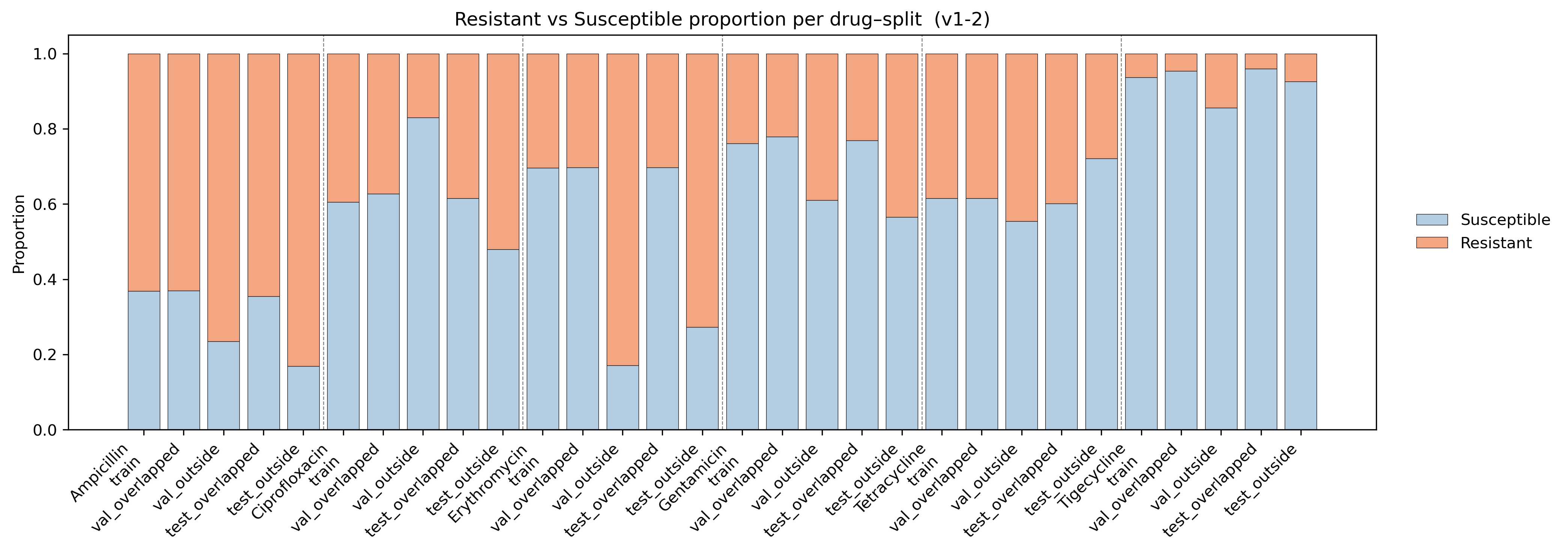}\\[6pt]
  \includegraphics[width=0.9\textwidth]{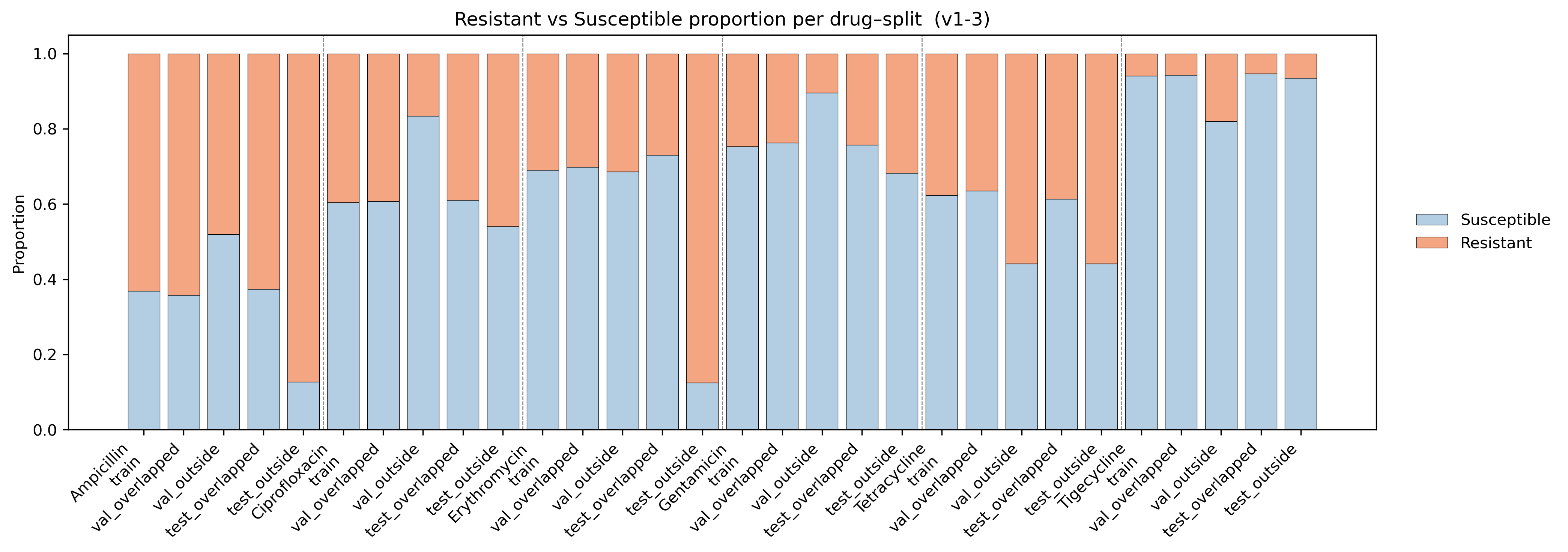}
  \caption{Label prevalence for the six retained antibiotics by partition across three replicates. Despite filtering for species diversity and sample size, natural class imbalance persists, ranging from 5\% to 65\% resistance rates. Replicate stability confirms reproducible split generation.}
  \label{fig:prop_drug_split}
\end{figure}

Having applied the retention criteria, we examine label prevalence within the six selected targets. Figure~\ref{fig:prop_drug_split} shows positive class rates across the five evaluation partitions for three independent replicates. While the filtering ensures adequate species diversity and sample sizes, inherent class imbalance remains—a natural characteristic of antimicrobial resistance data. Positive class prevalence ranges from approximately 5\% (Tigecycline) to 65\% (Ampicillin), but remains stable across replicates, confirming that our split procedure yields reproducible label distributions.

\section{Split Design and Purposeful Shift}

The split strategy is designed to disentangle memorization from generalization. Training and ``overlapped'' partitions share category support; ``outside'' partitions contain entirely unseen categories during fitting. Hyperparameters are tuned against \textit{val\_outside} to align model selection with the intended deployment regime. Final reporting averages metrics across three independently sampled replicates of the five-way split. This design reduces the sensitivity of results to a single random draw and approximates the practical setting in which new categories appear after model development.

\section{Quantifying Shift and Guarding Against Leakage}

We verify that no held-out category from \textit{outside} appears in \textit{train} or the overlapped sets. These checks ensure that performance on \textit{test\_outside} reflects genuine cross-category transfer rather than subtle contamination or degenerate baselines.

\section{Summary}

This chapter described the construction of a rigorous dataset for evaluating cross-species AMR prediction. Starting from \num{89451} bacterial genomes spanning 47 antimicrobial targets, we applied strict filtering criteria to retain six antibiotics with sufficient species diversity (minimum 5 species), sample sizes (minimum 100 resistant samples per partition), and total coverage (minimum 15,000 samples). The resulting dataset balances statistical power with taxonomic heterogeneity across ciprofloxacin, gentamicin, tetracycline, ampicillin, erythromycin, and tigecycline.

The five-way partitioning strategy---training, \textit{val\_overlapped}, \textit{val\_outside}, \textit{test\_overlapped}, and \textit{test\_outside}---creates a controlled distribution shift by holding out entire species from training. This design distinguishes true cross-species generalization from within-species memorization, with hyperparameter selection performed on \textit{val\_outside} to align model optimization with deployment conditions. Three independent replicates reduce sensitivity to partition artifacts.

With this carefully designed corpus in place, we can now evaluate whether foundation model representations truly generalize beyond their training distribution. Chapter~\ref{chap:kover} begins this evaluation by establishing baseline performance using traditional k-mer based methods, revealing the limitations of discrete sequence features for cross-species prediction.

%% file: chapter4.tex
\chapter{Rule-Based AMR Prediction: Benchmarking Kover's Generalization}
\label{chap:kover}

\section{Introduction}

Rule-based learners remain attractive for antimicrobial resistance (AMR) prediction because they produce human-readable rules that can be inspected against known mechanisms. Kover~\cite{Drouin2016} implements Set Covering Machines (SCM) and decision trees on binary k-mer features and has reported strong in-distribution performance~\cite{Hu2024}. This chapter examines whether such models transfer reliably across species, which is the regime most relevant to deployment.

Across six antibiotics under phylogenetically structured splits, we observe consistent cross-species degradation. Same-species performance is often stable, but accuracy drops sharply when evaluated on held-out species, with the magnitude depending on the drug. Instability is pronounced for ampicillin, where \texttt{val\_outside} F1 ranges from 0.03 to 0.87 across runs; tigecycline collapses under extreme class imbalance and predicts the majority class; erythromycin shows partition-sensitive behavior, occasionally achieving high F1 on validation while collapsing in specificity on test. These results motivate the context-preserving representations in Chapter~\ref{chap:amr_prediction}, which achieve substantially higher cross-species performance.

\section{Background and Rationale}

SCM constructs a sparse logical rule over literals derived from k-mer presence/absence. Given binary features $\{f_i\}_{i=1}^m$ and literals $\ell_i \in \{f_i,\neg f_i\}$, the prediction is an AND or OR over a small set $S$ selected to minimize empirical error under a budget on $|S|$. The appeal is interpretability and fast inference once trained. In this study we restrict Kover to its canonical binary classifier, SCM, to keep the comparison focused; decision trees in Kover are primarily used for multi-output tasks and add little for our binary setting.

\section{Experimental Setup}

We use the same phylogenetically structured partitions as elsewhere in the thesis. Training and the “overlapped’’ validations/tests draw from species observed during fitting; the “outside’’ validations/tests consist of species not seen during training. Hyperparameters are tuned on \texttt{val\_outside} to align model selection with the intended deployment shift rather than same-species fit. We grid-search the SCM complexity  
\[
p \in \{\,0.10,\;0.178,\;0.316,\;0.562,\;1.0,\;1.778,\;3.162,\;5.623,\;10.0,\;999999.0\,\}
\]
and rule type (AND/OR), select the configuration that maximizes F1 or balanced accuracy on \texttt{val\_outside}, and report all metrics on the held-out tests. We use $k=31$, the standard setting for Kover and the benchmark we compare against. Each experiment is repeated across three independently sampled split replicates to reduce sensitivity to a single partition.

\section{Results}

\begin{figure}[h]
\centering
\includegraphics[width=\textwidth]{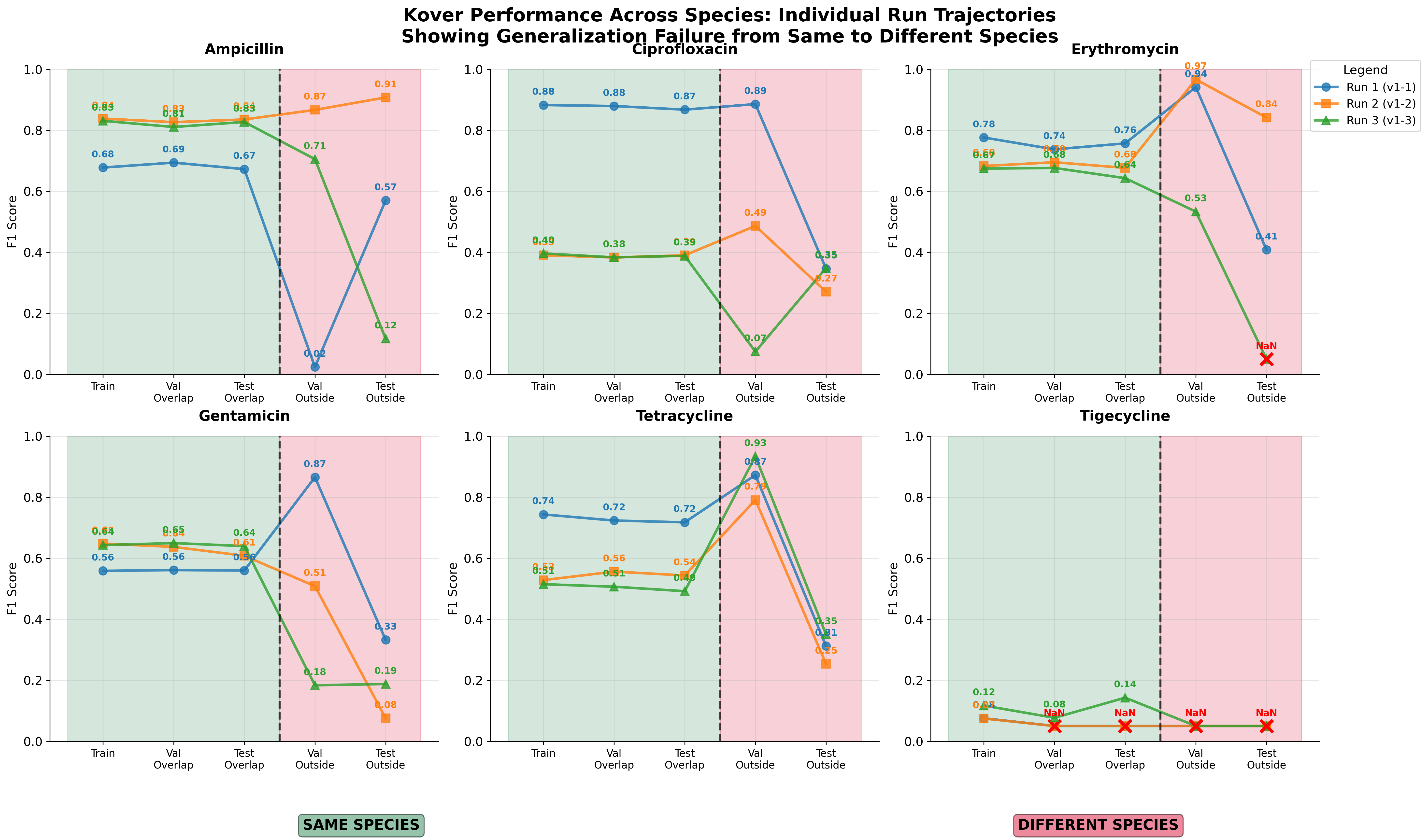}
\caption{\textbf{Cross-species degradation in Kover.} F1 across five partitions for six antibiotics (three runs where available). Green bands indicate same-species evaluation; pink bands indicate cross-species evaluation. The dashed line marks the transition to \texttt{val\_outside}. Degradation and variance are drug dependent; tigecycline fails under extreme imbalance.}
\label{fig:kover_trajectory}
\end{figure}

Figure~\ref{fig:kover_trajectory} summarizes F1 across partitions for six antibiotics. The pattern is consistent: performance within species is comparatively stable, while cross-species evaluation induces a drop whose size depends on the drug and replicate. Ampicillin shows the widest spread, with some runs discovering transferable motifs and others latching onto lineage proxies. Tigecycline illustrates a different failure mode: under extreme class imbalance (approximately 95:5 susceptible:resistant ratio), the model defaults to the susceptible majority. On several out-of-distribution splits it predicts no positives, making precision/recall undefined (denominator \(TP{+}FP=0\) and sometimes \(TP{+}FN=0\)), so $F_1$ is reported as NaN on \texttt{val\_outside} and \texttt{test\_outside}.

 Erythromycin alternates between apparently strong validation results and near 0 specificity on test, revealing partition sensitivity that same-species tuning would not surface. Ciprofloxacin, gentamicin, and tetracycline show substantial cross-species degradation, with magnitude varying by replicate.

Table~\ref{tab:species_boundary} summarizes the F1 ranges from 
Figure~\ref{fig:kover_trajectory}. The key signal is the between-run spread on \texttt{val\_outside} and \texttt{test\_outside}, particularly for ampicillin and erythromycin, which indicates sensitivity to which species are held out and to which literals the model selects under the budget constraint.

\begin{table}[h]
\centering
\small
\begin{tabular}{lcccc}
\toprule
\textbf{Antibiotic} & \textbf{Train F1} & \textbf{Val Outside F1} & \textbf{Test Outside F1} & \textbf{Run variance} \\
\midrule
Ampicillin & 0.68–0.84 & 0.02–0.87 & 0.12–0.91 & High/Extreme \\
Ciprofloxacin & 0.39–0.88 & 0.07–0.89 & 0.27–0.35 & High \\
Erythromycin & 0.67–0.78 & 0.53–0.97 & NaN–0.84 & High \\
Gentamicin & 0.56–0.65 & 0.18–0.87 & 0.08–0.33 & High \\
Tetracycline & 0.51–0.74 & 0.79–0.93 & 0.25–0.35 & Moderate \\
Tigecycline & 0.07–0.12 & NaN & NaN & Complete failure \\
\bottomrule
\end{tabular}
\caption{Approximate F1 ranges from Figure~\ref{fig:kover_trajectory}. Same-species performance is steady; cross-species results degrade and vary substantially across runs.}

\label{tab:species_boundary}
\end{table}
\footnotetext{\emph{NaN} indicates degenerate runs in which the classifier predicted no positives or the split contained no positive labels, yielding undefined precision or recall (i.e., division by zero in $TP{+}FP$ or $TP{+}FN$) and thus an undefined F1.}

\section{Analysis}

Two factors explain the observed failures. First, binary k-mer features entangle causal determinants (e.g., gene cassettes or specific mutations) with lineage background (e.g., GC content, codon bias, local genome organization). Within species, both correlate with phenotype; across species, only causal motifs transfer. A sparse logical rule may capture either, and when it leans on lineage proxies, cross-species accuracy collapses. Second, extreme class imbalance starves SCM of informative covering literals, pushing it toward majority-class predictions. Tigecycline is an instance of this problem and illustrates a practical limitation for low-prevalence resistance.

These effects persist even with OOD-aware tuning. Selecting hyperparameters on \texttt{val\_outside} reduces some overfitting to same-species quirks but cannot remove the ambiguity in what a short logical rule encodes under budget pressure. The between-run spread on ampicillin is a direct manifestation: some runs include transferable motifs, others do not, and the difference is invisible if tuning is done on \texttt{val\_overlapped} alone.

\section{Implications}

Rule-based models remain useful within known lineages, where their transparency is a virtue and inference is fast. They are less suitable for cross-species prediction, for highly imbalanced drugs that require sensitivity to rare positives, or for settings where robustness under shift is the primary objective. For cross-species deployment, representations that preserve sequence context and attenuate lineage background, rather than discrete k-mer presence alone, are required. Chapter~\ref{chap:amr_prediction} develop and evaluate such representations.

\section{Conclusion}

Kover achieves reasonable accuracy in-distribution but does not generalize reliably across species. Degradation at the species boundary, instability across runs, imbalance-driven failure, and partition sensitivity indicate that sparse k-mer rules often encode lineage rather than causal signal. Interpretability is valuable, but for cross-species clinical use, robustness is paramount. In the remainder of this thesis we therefore pivot to foundation-model representations designed to transfer across taxa.

%% file: chapter5.tex
\chapter{Diagnostic-Driven Layer Selection in Genomic Foundation Models}
\label{chap:evo_embedding}

\section{Introduction}

Extracting embeddings from the final layer is common practice, but often suboptimal for transfer learning. Transfer depends on extraction depth: mid-layers tend to be the most broadly useful, whereas the final layers are specialized to the pretraining objective and thus travel poorly~\cite{tenney2019bert,rogers2020primer,yosinski2014transfer,raghu2019transfusion,kornblith2019similarity}. The representation geometry also shifts with depth: upper layers become increasingly anisotropic, concentrating variance into a few directions and distorting neighborhood structure~\cite{ethayarajh2019contextual}. A practical third factor is numeric precision: modern inference commonly uses \texttt{bfloat16} (hereafter \texttt{bf16}); with only seven mantissa bits, large activations increase the unit-in-the-last-place (ULP) scale, perturbing angles, singular spectra, and any method that relies on cosine similarity~\cite{micikevicius2017mixed}.

This chapter turns these observations into a rule one can actually use with genomic foundation models: \emph{select the deepest stable layer}. Stability means three things measured under the model’s native mixed-precision numerics: activation scale remains moderate, angular geometry remains rich, and behavior is reproducible across random seeds. On \emph{Evo-1-8k-base} (32 blocks, width 4096), a single diagnostic sweep reveals a sharp stability boundary at Layer~11 (L11). Layer~10 (L10) sits immediately before that boundary and emerges as the deepest stable layer. All downstream experiments in this thesis therefore extract Evo embeddings at L10.

\section{Experimental Setup}

Our diagnostics are designed to separate biology from numerics and to observe the model \emph{as it computes}. From our bacterial collection we randomly select fifty genomes and, from each, six non-overlapping windows of \num{4000} tokens. In parallel we generate synthetic sequences that exercise the same numeric pathways while carrying no semantics: homopolymers, periodic repeats (period 10/100), cyclic shifts of a fixed base string, reversal, a full random permutation, and i.i.d.\ DNA. The depth-dependent patterns we report appear with the same character on these structureless controls, indicating that the behaviors are properties of the model’s computation (and mixed-precision numerics), not of genomic content.
Evo is evaluated under its official \texttt{bf16} policy (parameters in \texttt{bfloat16}, Hyena poles/residues in \texttt{float32}). We apply strict padding masks and check per-layer pad energy. Hidden states are materialized once per layer during a single forward pass, immediately cast to \texttt{float32} for statistics, logged, and freed; all curves aggregate ten independent seeds.

\section{Diagnostics}

We track five categories of diagnostic metrics designed to expose specific failure modes under mixed precision.

\emph{Activation tails.} For each depth $\ell$ we compute absolute activations for each token $\lvert h_\ell\rvert$ and record maxima and tail percentiles (p95, p99, p99.9, p99.99). In \texttt{bf16}, heavier tails imply larger relative rounding and unstable angles.

\emph{Standardized extremes.} After token-wise standardization $Z_\ell=(H_\ell-\mu_\ell)/\sigma_\ell$, we log $\max \lvert Z_\ell\rvert$ and the fraction of coordinates with $\lvert Z_\ell\rvert>6$. If only the scale changed, these would stay flat.

\emph{Energy concentration across tokens.} We monitor the mean, variance, and maximum of token $\ell_2$ norms and the mean/maximum of the largest $k$ token norms. A sudden increase indicates that a small subset of tokens dominates numerically, making downstream pooling and distance-based methods unstable.

\emph{Where extremes occur.} We record the most-activated coordinate index ($\operatorname*{argmax}_{\text{dim}}$) and the most-activated position ($\operatorname*{argmax}_{\text{token}}$). A narrowing $\operatorname*{argmax}_{\text{dim}}$ histogram is a direct footprint of anisotropy; a persistent first-position mode in $\operatorname*{argmax}_{\text{token}}$ is the hallmark of a sink at the beginning-of-sequence token.

\emph{Geometry.} Isotropy $\mathrm{Iso}(H)=1-\frac{1}{n}\sum_{i=1}^n\big|\langle \widehat{h}_i,\overline{\widehat{h}}\rangle\big|$, where $\widehat{h}_i$ denotes the $i$-th row-normalized embedding and $\overline{\widehat{h}}=\frac{1}{n}\sum_{i=1}^n\widehat{h}_i$ is their mean (higher isotropy indicates more angularly uniform representations); effective rank is computed from the singular-value ``distribution'' $p_i=\sigma_i/\sum_j \sigma_j$ with $\mathrm{erank}=\exp\!\big(H(p)\big)$. Both are reported with cross-seed variability.

\section{Results: A Stability Boundary at L11}

\begin{figure}[t]
  \centering
  \includegraphics[width=\textwidth]{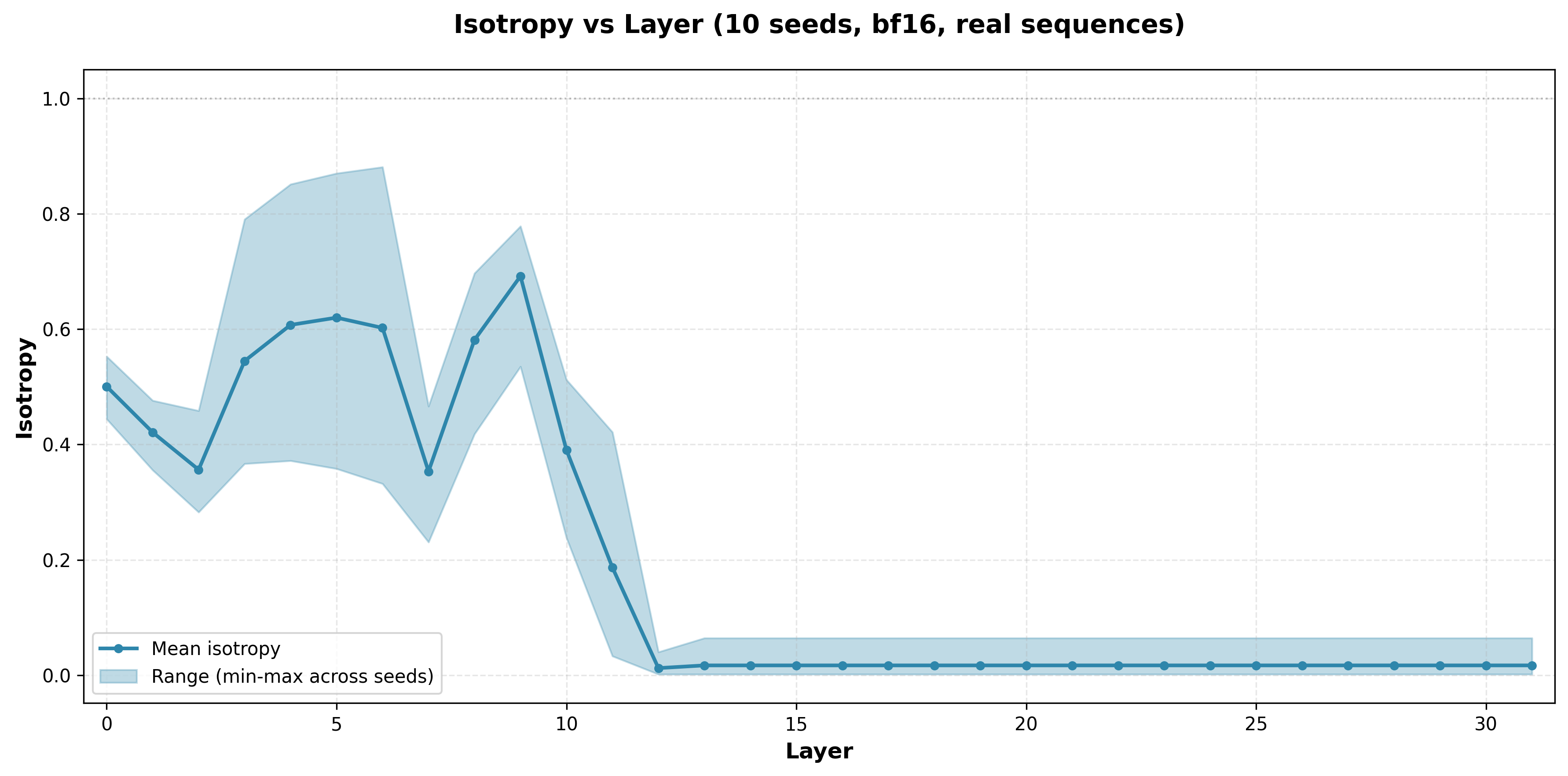}
  \caption{\textbf{Isotropy by depth.} Angular diversity increases through mid-layers, peaks at L9–L10, and collapses at L11 (ten seeds; min–max bands).}
  \label{fig:isotropy-depth}
\end{figure}

\begin{figure}[t]
  \centering
  \includegraphics[width=\textwidth]{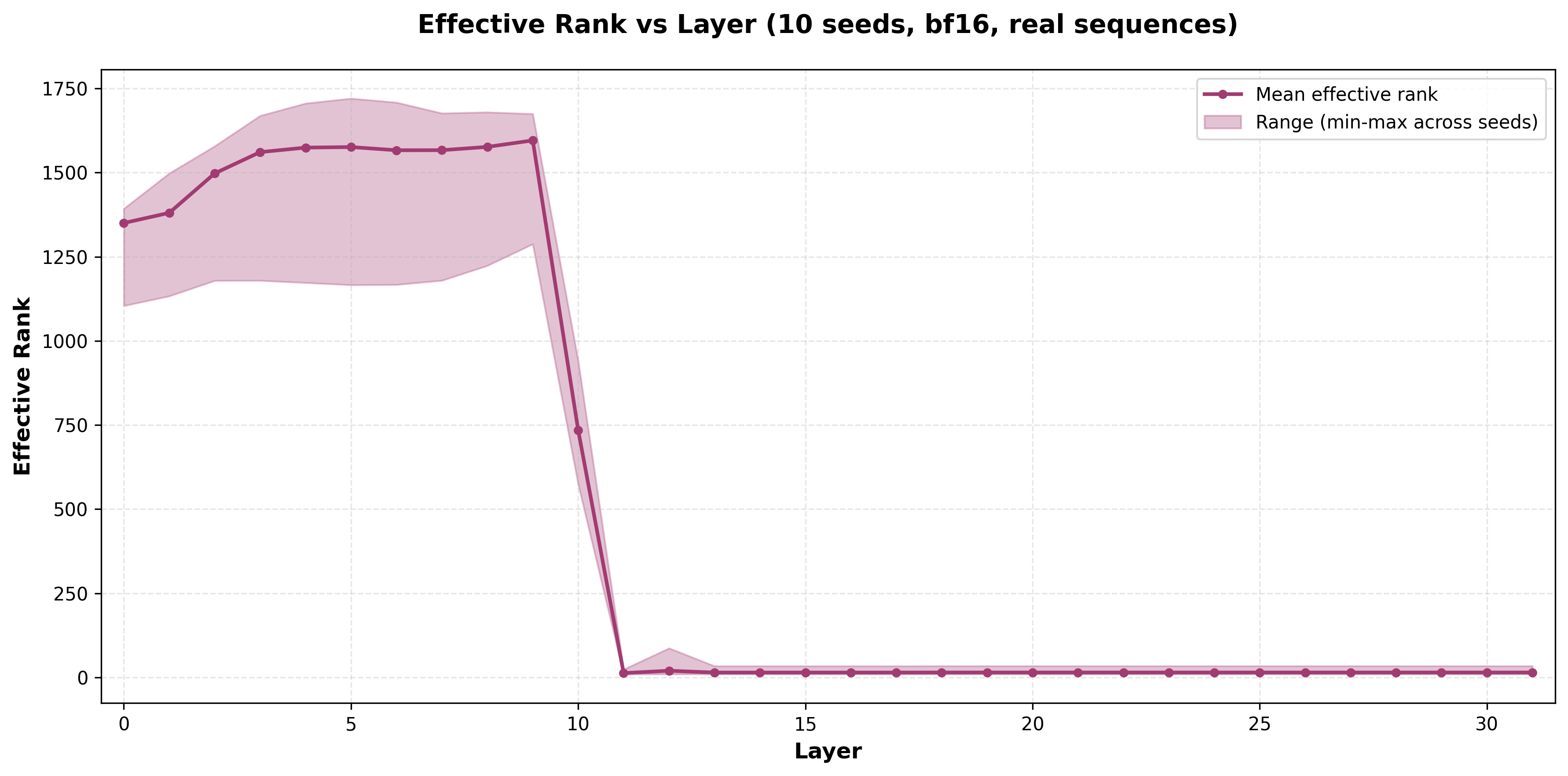}
  \caption{\textbf{Effective rank by depth.} The singular spectrum broadens until L9–L10 and compresses at L11, signaling entry into a low-dimensional subspace.}
  \label{fig:erank-depth}
\end{figure}

Two global views surface immediately. Isotropy as a function of depth (Figure~\ref{fig:isotropy-depth}) rises through the model, peaks around L9–L10, and then collapses at L11, remaining low thereafter. Effective rank (Figure~\ref{fig:erank-depth}) mirrors this story: the spectrum broadens to roughly \num{1600} effective dimensions by L9–L10 and then compresses sharply at L11. Together they indicate a point where geometry stops improving and starts degenerating.

\begin{figure}[t]
  \centering
  \includegraphics[width=\textwidth]{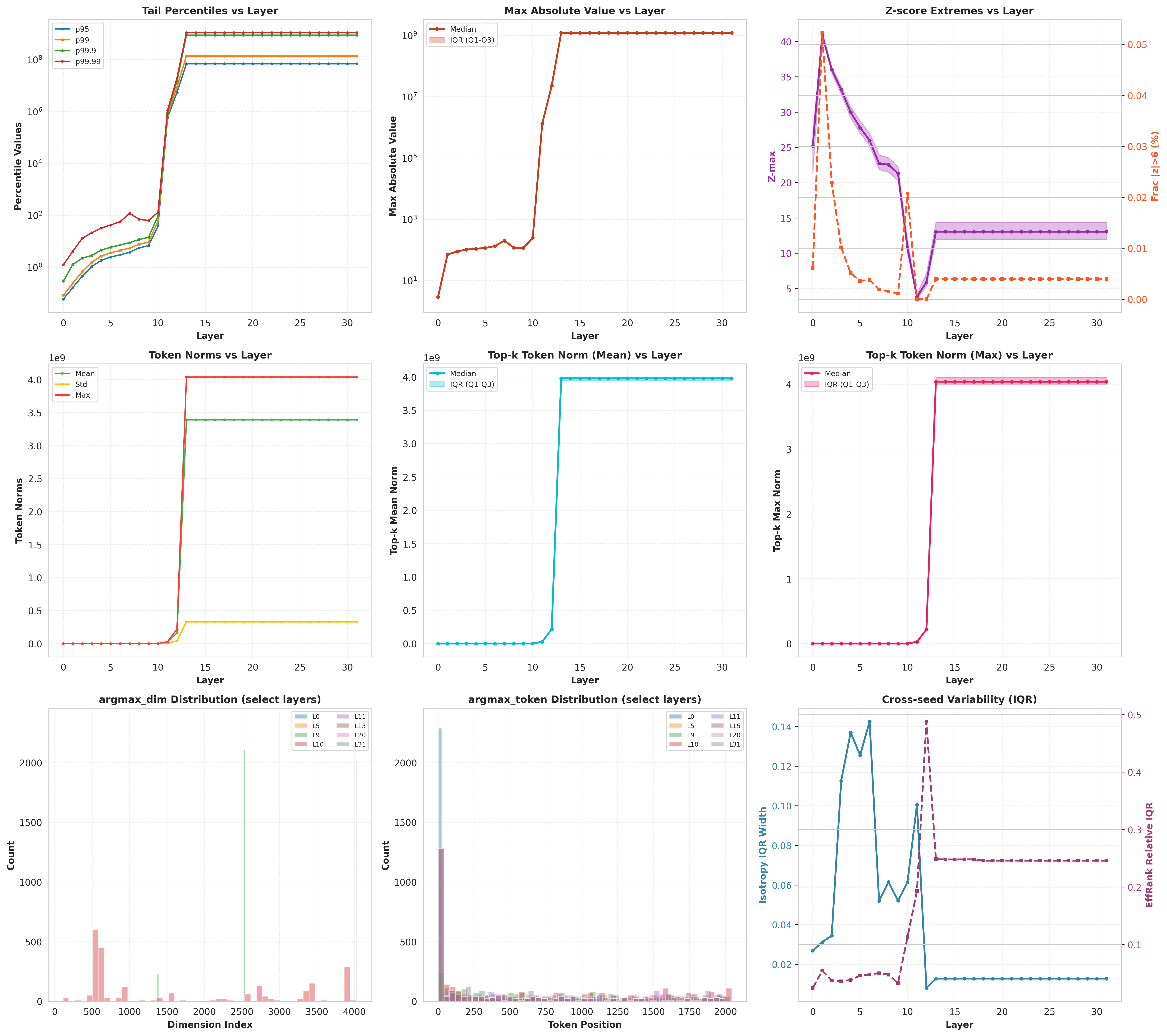}
  \caption{\textbf{Layer diagnostics under native \texttt{bf16} (ten seeds; medians with uncertainty bands).}
  \emph{Top:} heavy tails and maxima remain moderate through L10 and jump at L11; standardized extremes corroborate a shape change.
  \emph{Middle:} token-norm concentration remains contained until L10; at L11, top-$k$ tokens dominate.
  \emph{Bottom:} deep-layer anisotropy (narrow $\operatorname*{argmax}_{\text{dim}}$), a strengthening first-token sink in $\operatorname*{argmax}_{\text{token}}$, and a sharp rise in cross-seed variability.}
  \label{fig:layer_diag_grid}
\end{figure}

The nine-panel grid in Figure~\ref{fig:layer_diag_grid} reveals the mechanism. The top row shows that activation tails lift steadily and then jump in unison at L11; per-layer maxima follow the same step. After standardization, both $\max\lvert Z\rvert$ and the mass beyond six standard deviations remain negligible up to L10 and then increase—evidence of a distributional shape change that a simple rescale cannot explain. In the middle row, mean and variance of token norms continue their gradual rise, but the maximum token norm and the top-$k$ norms stay controlled through L10 and surge at L11: a small set of tokens takes over the numeric budget. The bottom row localizes the effect. The $\operatorname*{argmax}_{\text{dim}}$ histogram narrows in deeper layers, consistent with depth-induced anisotropy~\cite{ethayarajh2019contextual}. The $\operatorname*{argmax}_{\text{token}}$ histogram shows a persistent first-token mode across depth, which strengthens dramatically after L11. Concomitantly, cross-seed variability (IQR and relative IQR) spikes for isotropy and effective rank. Even if a single run were to look acceptable, repeated trials would not agree; beyond L11, the representation is not just low-rank and anisotropic, it is also unreliable.

\section{Mechanism: Sinks, Compression, and Massive Residuals}

A recent theoretical account unifies two empirical puzzles in LLMs—attention sinks and compression valleys—as two faces of the same mechanism driven by massive residual activations~\cite{queipo2025sinks}. The theory predicts that when the first token develops extreme activation norms around middle depth, the representation must compress, lowering entropy and effective rank; attention then fixates on the sink. Our data follow this script precisely at L11. Activation tails and maxima step up; top-$k$ token norms surge; the first-token mode strengthens; isotropy collapses; and the singular spectrum compresses (Figures~\ref{fig:isotropy-depth}–\ref{fig:layer_diag_grid}). In the language of that work, Evo passes from an early \emph{mix} phase to a \emph{compress} phase; for embedding tasks, the best extraction point lies just before compression fully sets in.

\section{Why L10 Is the Right Choice}

Three lines of evidence converge. First, depth–transfer studies consistently favor middle layers for off-task generalization~\cite{tenney2019bert,rogers2020primer,yosinski2014transfer,raghu2019transfusion,kornblith2019similarity}; our isotropy and effective-rank peaks at L9–L10 are their geometric signature. Second, geometry itself degrades late: deep-layer anisotropy is visible in the narrowing $\operatorname*{argmax}_{\text{dim}}$, the rank collapse, and the persistent first-token mode. Third, numerics worsen exactly where geometry does: in \texttt{bf16} the ULP 
grows with magnitude, so the L11 increase in tails and maxima directly inflates 
angular error and perturbs SVDs~\cite{micikevicius2017mixed}. L10 is the deepest 
layer where transferability, geometry, and \texttt{bf16} stability hold jointly.

\section{Robustness}

Batch and window order do not matter. Synthetic controls—homopolymers, periodic repeats, permutations—exhibit the same L11 jump in tails and the same compression of geometry, indicating that the phenomenon is a property of Evo’s residual amplification under \texttt{bf16} rather than an artifact of genomic content. Varying the genome list and resampling windows keep the boundary within the uncertainty bands shown.

\section{Implications for AMR and Systems}

For downstream AMR, the difference is tangible. L10 embeddings retain angular diversity and broader spectra, which improves the conditioning of PCA and sparse random projection (SRP), stabilizes cosine kernels, and makes MiniRocket features more informative by preserving local statistics rather than letting a handful of saturated tokens dominate. Numerically, moderate scales at L10 reduce \texttt{bf16} rounding noise; seed-to-seed variance tightens and training curves become reproducible. Operationally, fixing the extraction depth enables a streaming extractor that 
never stages full layer stacks, which in our implementation cut GPU memory 
requirements by roughly an order of magnitude.

\section{Limitations}

The diagnostics are deliberately one-pass: we do not ablate internal submodules or retrain layers, so we do not claim causal attribution to a specific component. The choice of L10 is also task-aware: tasks tightly aligned with Evo’s pretraining may prefer deeper layers. Our claim is narrower and sufficient for what follows: under native \texttt{bf16}, L10 is the deepest layer whose geometry and numerics are jointly stable across a mixed corpus; choosing it improves cross-species AMR in Chapter~\ref{chap:amr_prediction}.

\section{Conclusion}

A single diagnostic sweep is enough to locate a stability boundary in \emph{Evo-1-8k-base}. Activation tails, standardized extremes, token-norm concentration, first-token localization, isotropy, effective rank, and cross-seed variability all change phase at L11. Layer~10 lies just before that transition and is therefore the \emph{deepest stable layer}. This selection accords with depth–transfer evidence~\cite{tenney2019bert,rogers2020primer,yosinski2014transfer,raghu2019transfusion,kornblith2019similarity}, with depth-induced anisotropy~\cite{ethayarajh2019contextual}, with \texttt{bf16} numerics~\cite{micikevicius2017mixed}, and with the theoretical link between attention sinks and compression valleys~\cite{queipo2025sinks}. We adopt L10 uniformly in the remainder of this thesis.

%% file: chapter6.tex
\chapter{Cross-Species AMR Prediction via Local Pattern Preservation}
\label{chap:amr_prediction}

\section{Introduction}

This chapter studies antimicrobial resistance (AMR) prediction from Evo foundation model embeddings and asks a concrete question: can we improve cross-species generalization by treating embeddings as \emph{ordered multivariate signals} rather than as a single pooled vector?

Conventional aggregation pipelines work well when train and test species match, but they often degrade sharply when evaluated on new species. Global pooling emphasizes genome-wide composition, which tracks species identity more strongly than resistance phenotype. In contrast, many ampicillin resistance mechanisms are carried by \emph{modular, horizontally transferable} elements such as $\beta$-lactamase cassettes on plasmids, transposons, and integrons. These modules are local in the genome and potentially portable across taxa, suggesting that an aggregation scheme that preserves locality might surface cross-species AMR signatures more effectively.

We compare two aggregation strategies applied to Evo Layer~10 (L10) token embeddings. Both create compressed representations of genome-scale embeddings, but differ fundamentally in how they handle spatial structure. \textbf{Global Pooling} applies a fixed PCA projection (4{,}096 $\rightarrow$ 41 dimensions) to each token embedding, then computes six summary statistics across all tokens (mean, standard deviation, minimum, maximum, 25th and 75th percentiles) per PCA channel, yielding $41 \times 6 = 246$ features per genome. This approach discards positional information. \textbf{MiniRocket} instead maintains the ordered sequence of 41-dimensional token embeddings, processes overlapping token chunks with MiniRocket's fixed binary convolutions, and summarizes kernel responses using the proportion of positive values (PPV), producing $\approx 12{,}000$ features per genome that encode local pattern statistics while preserving spatial relationships.

Two empirical observations emerge. First, cross-species performance is highly inconsistent across splits: MiniRocket dramatically improves k-NN performance on \texttt{val\_outside} (MCC: 0.753 vs.\ 0.148), while Global Pooling dominates on \texttt{test\_outside} for most classifiers (e.g., LightGBM: 0.932 vs.\ 0.798). This split-dependent behavior initially appears contradictory—neither method universally dominates cross-species prediction. Under the mechanism-mix hypothesis, however, such reversals are plausible: different held-out species sets may differ in their dominant resistance mechanisms, so split-level aggregates can flip even when the underlying relationship between aggregation strategy and mechanism type is consistent.

This inconsistency motivated species-level analysis, which reveals the underlying pattern: performance depends on resistance mechanism, not phylogenetic distance alone. Species whose resistance is plausibly driven by local, horizontally transferable elements (e.g., plasmid-borne $\beta$-lactamases) show dramatic accuracy improvements with MiniRocket, while species with diffuse or chromosomal mechanisms show comparable or degraded performance. Split-level aggregates therefore reflect which mechanism profiles dominate each held-out species set, not fundamental method superiority.

Second, after the MiniRocket transform, simple $k$-nearest neighbor (k-NN) classifiers become top performers on \texttt{val\_outside} (MCC up to 0.753), suggesting that the transformed space clusters genomes in a way that is more consistent with shared resistance modules than with phylogeny. Neighbor analysis supports this interpretation: test genomes shift from selecting phylogenetically similar neighbors to selecting neighbors from a small set of training species. On same-species splits (\texttt{val\_overlapped}, \texttt{test\_overlapped}), both pipelines achieve similar performance, with the best model varying by metric and classifier. The rest of this chapter describes how the aggregation is implemented, how the experiment is designed, and how these changes in feature space geometry relate to phylogeny and species-level behavior.

\section{From Static Vectors to Signals}

\subsection{Rethinking Embedding Aggregation}

In most genomic foundation model workflows, embeddings are extracted from windows or full sequences and then collapsed to a single vector. DNABERT and the Nucleotide Transformer commonly use a CLS token, HyenaDNA uses an EOS token, and mean pooling across tokens often performs best in practice~\cite{Ji2021,Zhou2023,DallaTorre2023,nguyen2023hyenadna}.

Time series methods such as ROCKET~\cite{Dempster2020} and MiniRocket~\cite{Dempster2021} take the opposite perspective: they assume a multivariate signal and construct large banks of fixed convolutions to produce features for linear classifiers. Our approach combines these ideas. We treat the ordered sequence of Evo token embeddings as a multivariate signal indexed by genomic position and then apply MiniRocket as a deterministic pattern detector. To our knowledge, prior work has not combined large genomic foundation models with time series feature extractors in this way.

Genomic position is not temporal, but it is structured. Genes cluster into operons, mobile elements occupy contiguous loci, and regulatory elements show local patterns. Averaging across an entire chromosome can easily drown out sparse but important modules (for example, a 3~kb cassette on a 4~Mb genome). Preserving order and applying local pattern detectors allows us to retain these signals while still producing a fixed-length feature vector.

\subsection{Aggregation Strategies}

Evo produces a 4{,}096-dimensional embedding for each token. For I/O efficiency we store embeddings in non-overlapping $\sim$4~kb blocks, but all aggregation operates at the token level. Throughout this chapter, a ``token'' refers to Evo's native sequence token; in our setting, 4{,}000 tokens correspond approximately to a 4~kb genomic span. After PCA (4{,}096 $\rightarrow$ 41 per token), each genome becomes a length-$T$ sequence $\{\mathbf{z}_t\}_{t=1}^{T}$ with $\mathbf{z}_t \in \mathbb{R}^{41}$. The PCA projection is fit on training-set token embeddings only and then applied to all splits.

\paragraph{Global pooling.}
Global Pooling pools across tokens. For each of the 41 channels we compute mean, standard deviation, minimum, maximum, and the 25th and 75th percentiles across tokens, and then concatenate all channel-wise statistics into a 246-dimensional feature vector. This is efficient and stable but discards the spatial structure of the embedding stream.

\paragraph{MiniRocket local pattern preservation.}
MiniRocket instead views $\{\mathbf{z}_t\}$ as a multivariate signal. We process overlapping token chunks of length $L = 2048$ with stride $S = 1024$ (all in tokens). Within each chunk, MiniRocket applies $K = 2000$ fixed binary kernels of length~9 at log-spaced dilations. We use a simple multivariate reduction: for each kernel, responses are computed per channel and summed across randomly selected channels before PPV thresholding. This matches our implementation; channel subsets are sampled once per kernel with a fixed random seed, so the transform is deterministic aside from the one-time bias fit. 
PPVs are computed for each chunk and each kernel; for each kernel we then summarize the resulting PPV values across chunks using six statistics (mean, standard deviation, minimum, maximum, 25th and 75th percentiles), yielding $K \times 6 \approx 12{,}000$ features per genome.


We adapt MiniRocket to token embedding streams using Algorithms~\ref{alg:minirocket-fit}--\ref{alg:minirocket-transform}. The kernels and dilations are deterministic; only the kernel biases are fit once on training data and then reused. This preserves positional structure without introducing extra trainable parameters beyond the downstream classifier.

\begin{algorithm}[htbp]
\caption{MiniRocket-Fit (Token-Level)}
\label{alg:minirocket-fit}
\Input{Training streams $\{\mathbf{Z}^{(n)}\}_{n=1}^N$, each $\mathbf{Z}^{(n)}=\{\mathbf{z}^{(n)}_t\}_{t=1}^{T_n}$, $\mathbf{z}_t\in\mathbb{R}^{C}$ with $C{=}41$; chunk length $L{=}2048$ tokens, stride $S{=}1024$ tokens, kernel length $\ell{=}9$, kernels $K{=}2000$}
\Output{Kernel patterns $\{(\mathcal{I}^{+}_p,\mathcal{I}^{-}_p)\}_{p=1}^{K}$, dilations $\mathcal{D}$, counts $\{N_d\}$, biases $\mathbf{B}\in\mathbb{R}^{K}$, channel subsets $\{S_p\}_{p=1}^{K}$}
\BlankLine
Generate $K$ zero-sum binary patterns $\{(\mathcal{I}^{+}_p,\mathcal{I}^{-}_p)\}_{p=1}^{K}$ (three $+2$, six $-1$)\;
Choose log-spaced $\mathcal{D}\subset\{1,\dots,L{-}1\}$ and allocate $\{N_d\}$\;
$p \gets 1$\;
\For{$d\in\mathcal{D}$}{
  \For{$j=1$ \KwTo $N_d$}{
    Sample a training chunk $\mathbf{X}\in\mathbb{R}^{L\times C}$\;
    Sample a channel subset $S_p \subseteq \{1,\dots,C\}$ (fixed seed, deterministic)\;
    Compute responses
    $r_{\tau} = \sum_{c\in S_p}\!\left(2\!\!\sum_{i\in\mathcal{I}^{+}_p}\!X_{\tau+i d,c} - \sum_{i\in\mathcal{I}^{-}_p}\!X_{\tau+i d,c}\right)$
    over valid positions $\mathcal{U}$\;
    Set $\mathbf{B}[p]$ to a predetermined quantile (e.g., $q_{0.84}$, as in MiniRocket) of $\{r_{\tau}\}_{\tau \in \mathcal{U}}$\;
    $p \gets p{+}1$\;
  }
}
\KwRet{$(\{(\mathcal{I}^{+}_p,\mathcal{I}^{-}_p)\}, \mathcal{D}, \{N_d\}, \mathbf{B}, \{S_p\})$}\;
\end{algorithm}

\begin{algorithm}[htbp]
\caption{MiniRocket-Transform (Token-Level)}
\label{alg:minirocket-transform}
\Input{Stream $\mathbf{Z}=\{\mathbf{z}_t\}_{t=1}^{T}$, $\mathbf{z}_t\in\mathbb{R}^{41}$; $L{=}2048$, $S{=}1024$, $\ell{=}9$, kernel patterns $\{(\mathcal{I}^{+}_p,\mathcal{I}^{-}_p)\}$, dilations $\mathcal{D}$, counts $\{N_d\}$, biases $\mathbf{B}$, channel subsets $\{S_p\}_{p=1}^{K}$}
\Output{Feature vector $\mathbf{f}\in\mathbb{R}^{6K}$}
\BlankLine
Initialize lists $\{\mathcal{Y}_p\}_{p=1}^{K}$ as empty\;
\For{$s\in\{1,1{+}S,\dots\}$ with $s{+}L{-}1\le T$}{
  $\mathbf{X}\gets \mathbf{Z}[s{:}s{+}L{-}1]\in\mathbb{R}^{L\times 41}$; $p\gets 1$\;
  \For{$d\in\mathcal{D}$}{
    \For{$j=1$ \KwTo $N_d$}{
      Compute responses
      $r_{\tau} = \sum_{c\in S_p}\!\left(2\!\!\sum_{i\in\mathcal{I}^{+}_p}\!X_{\tau+i d,c} - \sum_{i\in\mathcal{I}^{-}_p}\!X_{\tau+i d,c}\right)$
      over valid positions $\mathcal{U}$\;
      $y \gets \frac{1}{|\mathcal{U}|}\sum_{\tau \in \mathcal{U}}\mathbf{1}\{r_{\tau} > \mathbf{B}[p]\}$ \tcp*{Proportion of Positive Values (PPV)}
      Append $y$ to $\mathcal{Y}_p$; $p\gets p + 1$\;
    }
  }
}
\For{$p=1$ \KwTo $K$}{
  $\mathbf{g}_p \gets \big(\mathrm{mean}(\mathcal{Y}_p),\,\mathrm{std}(\mathcal{Y}_p),\,\min(\mathcal{Y}_p),\,\max(\mathcal{Y}_p),\,q_{0.25}(\mathcal{Y}_p),\,q_{0.75}(\mathcal{Y}_p)\big)$\;
  $\mathbf{f}[6(p-1)+1:6p] \gets \mathbf{g}_p$\;
}
\KwRet{$\mathbf{f}\in\mathbb{R}^{6K}$} \tcp*{6 stats of PPVs across chunks per kernel}
\end{algorithm}


\paragraph{Implementation notes.}

Genome-level features are six summary statistics of PPVs across all chunks per kernel (mean, standard deviation, minimum, maximum, 25th percentile, and 75th percentile), resulting in $6K \approx 12{,}000$ features per genome.

\section{Experimental Design}

We evaluate bacterial 3{,}388 genomes spanning \num{126} species with experimentally validated ampicillin resistance labels. We treat resistance as the positive class; across splits, resistance prevalence ranges from 64\% to 71\%. The data are partitioned to probe different generalization regimes:

\begin{itemize}
    \item \textbf{Train} ($n = 2000$): genomes from a fixed set of species.
    \item \textbf{\texttt{val\_overlapped}} ($n = 336$): genomes from the same species as \textbf{Train}, with no strain overlap.
    \item \textbf{\texttt{val\_outside}} ($n = 260$): genomes from species not present in \textbf{Train}.
    \item \textbf{\texttt{test\_overlapped}} ($n = 394$): genomes from the same species as \textbf{Train}.
    \item \textbf{\texttt{test\_outside}} ($n = 398$): genomes from species not present in \textbf{Train} or \textbf{\texttt{val\_outside}}.
\end{itemize}

\noindent\textbf{Subsampling for embedding extraction.}
Because extracting Evo embeddings for every genome in the full dataset is computationally expensive, we compute embeddings on a subsample of the genomes used in the corresponding Kover splits. We subsample within each split using stratified sampling proportional to species frequency, while ensuring that every species present in the split remains represented. To verify that this subsampling does not materially change conclusions, we perform incremental sampling: we progressively add additional random batches and re-evaluate performance, observing stable metrics as sample size increases. We therefore treat the embedding-based results as comparable to the Kover baselines under the same split definitions.

Because the species are phylogenetically diverse, \texttt{val\_outside} and \texttt{test\_outside} contain both species from genera represented in training and species from entirely new genera. We report Matthews Correlation Coefficient (MCC), F1, AUROC, and AUPRC, but rely on MCC and AUPRC for most conclusions, since F1 and AUROC can be optimistic under class imbalance.

The Global Pooling and MiniRocket feature pipelines described above are paired with a common panel of classifiers: cosine k-NN with $k \in \{5, 10, 15, 20\}$, ExtraTrees, Random Forest, LightGBM, XGBoost, linear and RBF SVMs, logistic regression variants, and small and medium multi-layer perceptrons.

\section{Results: Changes in Feature Space Geometry}

\subsection{Overall Performance}
\label{sec:overall_performance}

\begin{figure}[htbp]
\centering
\includegraphics[width=\textwidth]{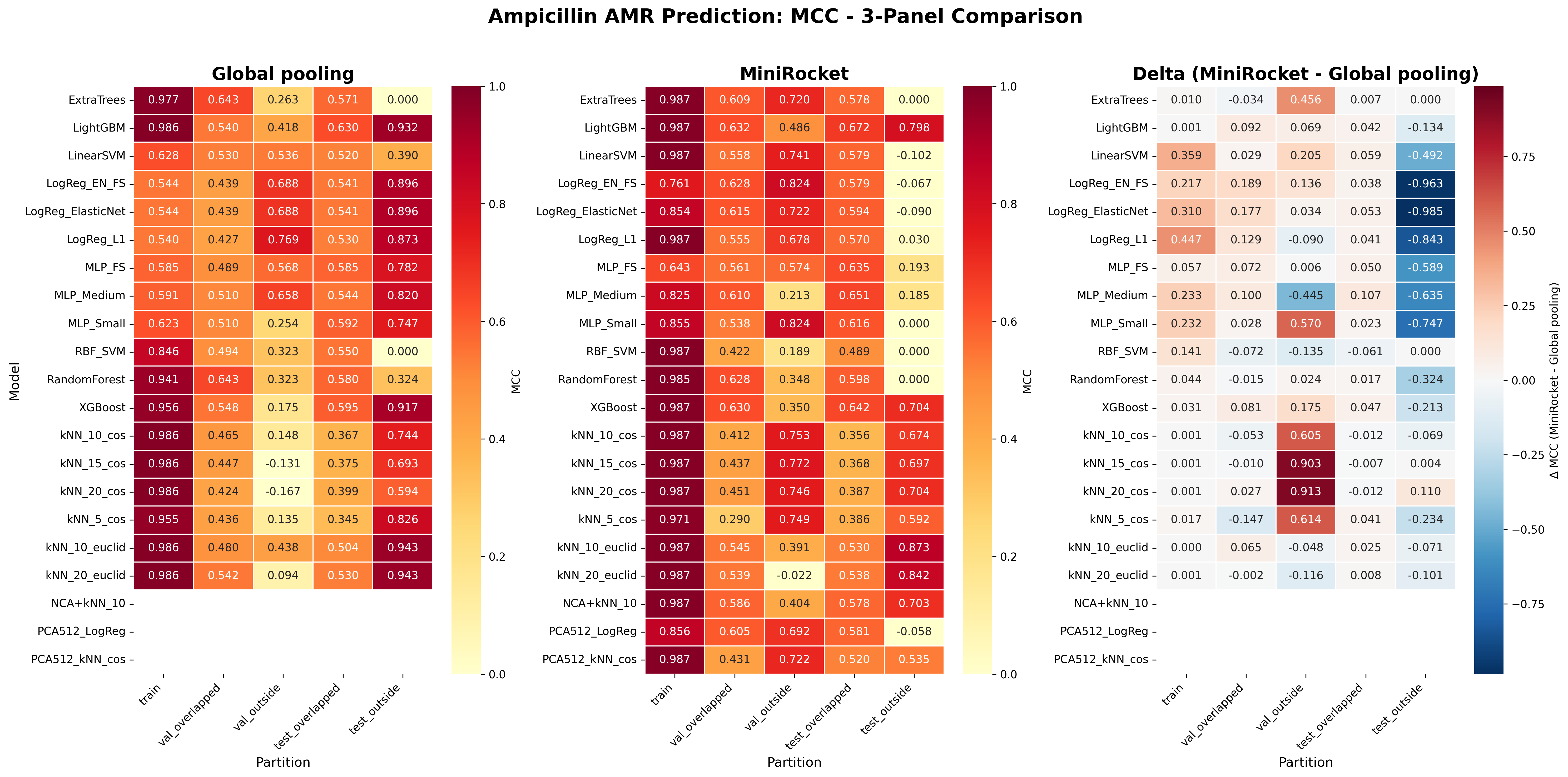}
\caption{\textbf{Ampicillin MCC across models and partitions.} Global Pooling (left) and MiniRocket (middle) MCC for each classifier and split; the right panel shows MiniRocket--Global Pooling deltas. MiniRocket and Global Pooling are similar on same-species splits (\texttt{val\_overlapped}, \texttt{test\_overlapped}). On \texttt{val\_outside}, MiniRocket yields large MCC gains, especially for k-NN. On \texttt{test\_outside}, Global Pooling outperforms MiniRocket for most classifiers. These split-dependent patterns motivated species-level analysis to identify which resistance mechanisms benefit from local pattern preservation.}
\label{fig:mcc_comparison}
\end{figure}

Figure~\ref{fig:mcc_comparison} summarizes MCC across all models and partitions. Cross-species performance shows stark inconsistencies across splits. On \texttt{val\_outside}, MiniRocket with k-NN achieves MCC $= 0.753$ (vs.\ $0.148$ for Global Pooling k-NN), with similar large gains for other k-NN variants ($\Delta$MCC: $+0.605$ to $+0.913$). On \texttt{test\_outside}, the pattern reverses: Global Pooling dominates for most classifiers (LightGBM: 0.932 vs.\ 0.798; XGBoost: 0.917 vs.\ 0.704), though k-NN results remain mixed. These split-dependent patterns suggest that neither method universally dominates, motivating species-level analysis.

On same-species splits (\texttt{val\_overlapped}, \texttt{test\_overlapped}), differences are much smaller. Depending on the classifier, either pipeline can be slightly ahead. Global Pooling models achieve very high MCC on the training split, reflecting strong within-species fit and the fact that global pooling is well aligned with within-species prediction. Rather than reflecting fundamental method differences, these split-level aggregates likely reflect which resistance mechanisms dominate each held-out species set—a hypothesis we test through species-level neighbor analysis in Section~\ref{sec:neighbor_analysis}.

\begin{figure}[htbp]
\centering
\includegraphics[width=\textwidth]{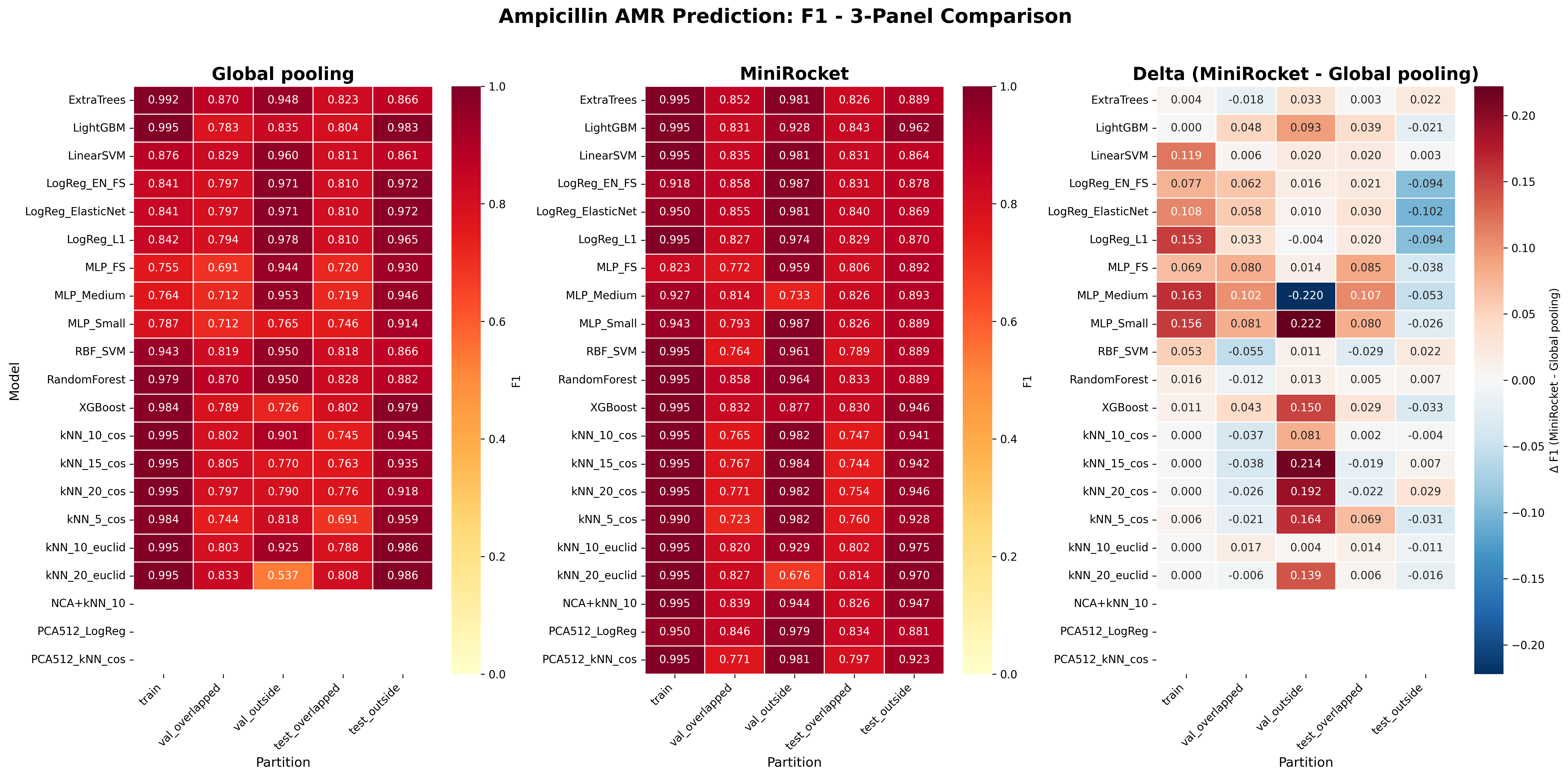}
\caption{\textbf{Ampicillin F1 across models and partitions.} Both pipelines achieve high F1 scores on all splits, but with different best-performing models, indicating that the feature spaces after the two transformations are fundamentally different.}
\label{fig:f1_comparison}
\end{figure}

Figure~\ref{fig:f1_comparison} shows that both pipelines attain high F1 scores (often 0.90--0.98) on all splits. If we considered F1 alone, the two approaches would appear similarly strong, including on cross-species evaluation. However, the MCC heatmap reveals that MiniRocket achieves substantially more balanced predictions on \texttt{val\_outside}, particularly by improving balanced performance as reflected in MCC for k-NN classifiers, while Global Pooling shows poor class balance as phylogenetic distance increases. On \texttt{test\_outside}, the difference is less pronounced, with Global Pooling maintaining reasonable balance for several classifiers.

\begin{figure}[htbp]
\centering
\includegraphics[width=\textwidth]{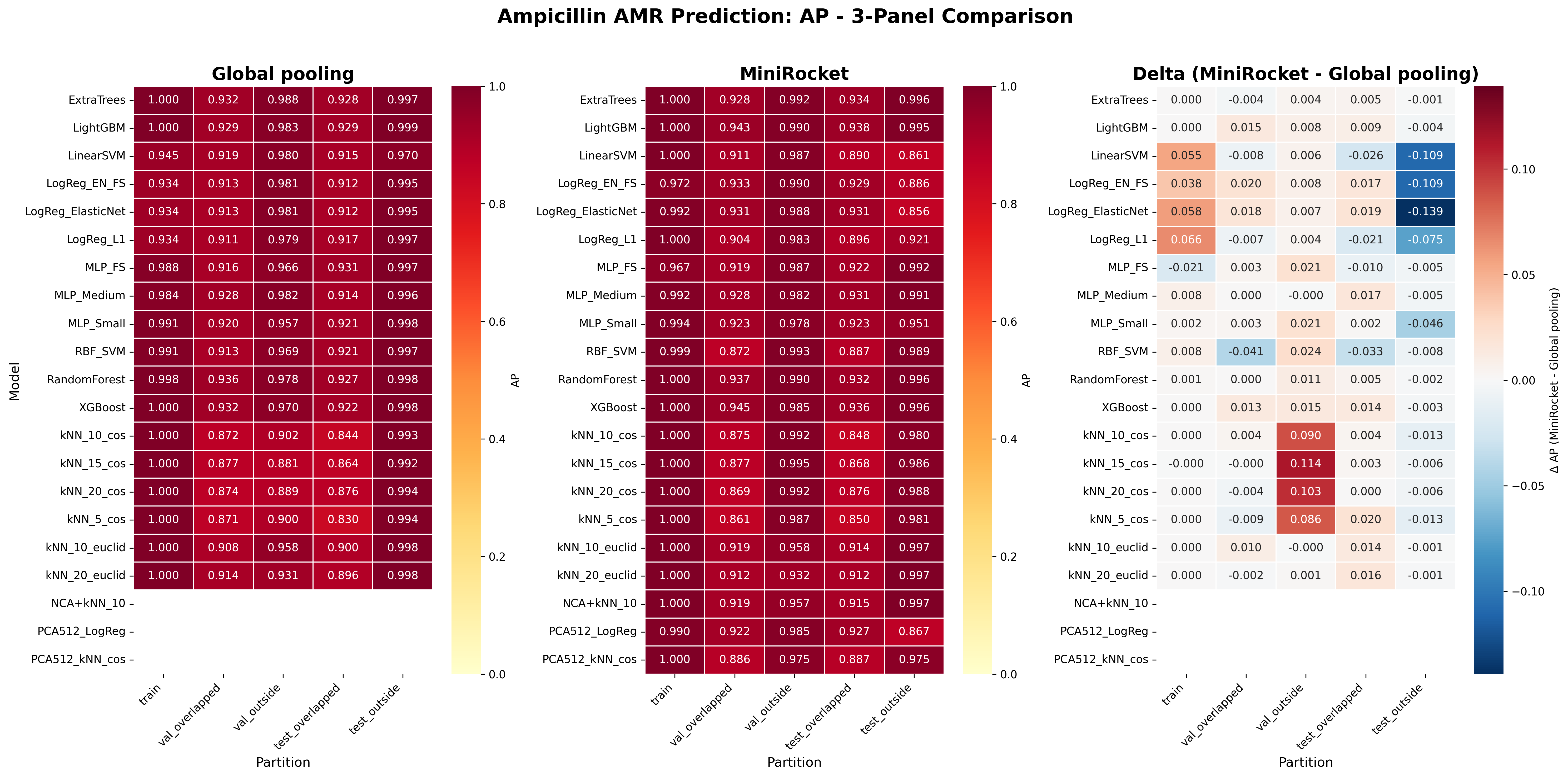}
\caption{\textbf{Ampicillin AUPRC across models and partitions.} AUPRC gains on \texttt{val\_outside} closely track MCC gains, indicating that MiniRocket improves the underlying precision--recall structure for k-NN and several other classifiers. On \texttt{test\_outside}, both approaches achieve high AUPRC, with smaller differences between pipelines.}
\label{fig:auprc_comparison}
\end{figure}

\begin{figure}[htbp]
\centering
\includegraphics[width=\textwidth]{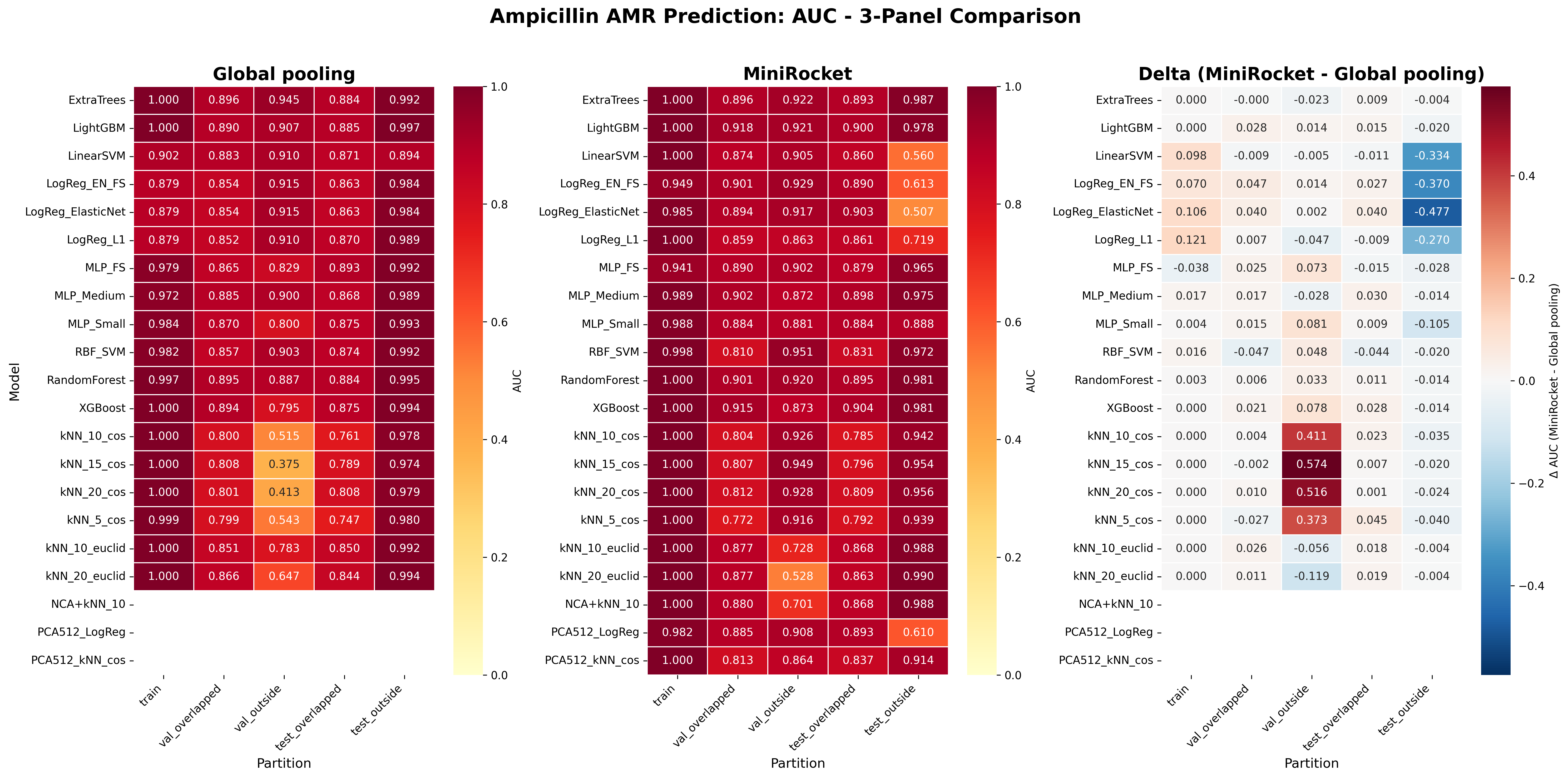}
\caption{\textbf{Ampicillin AUROC across models and partitions.} AUROC differences are smaller than MCC and AUPRC differences, as expected under class imbalance. For k-NN on \texttt{val\_outside}, MiniRocket increases AUROC from 0.515 to 0.926. On \texttt{test\_outside}, several linear and tree-based models achieve high AUROC with either pipeline, though some logistic regression models show lower AUROC with MiniRocket.}
\label{fig:auroc_comparison}
\end{figure}

AUPRC (Figure~\ref{fig:auprc_comparison}) on \texttt{val\_outside} closely tracks MCC: MiniRocket substantially improves AUPRC for k-NN (from 0.902 to 0.992) and maintains high AUPRC for most other classifiers. On same-species splits and \texttt{test\_outside}, both pipelines achieve high AUPRC with smaller differences. AUROC (Figure~\ref{fig:auroc_comparison}) is more conservative overall; many classifiers achieve high AUROC with either pipeline. The most dramatic AUROC improvement appears for k-NN on \texttt{val\_outside} (0.515 to 0.926), while on \texttt{test\_outside}, both pipelines yield high AUROC for most models, though some logistic regression variants show reduced AUROC with MiniRocket. Taken together, MCC and AUPRC support the view that performance depends on held-out species composition rather than split identity per se. We therefore turn to species-level neighbor analysis to identify which genomes benefit from local pattern preservation and which do not.

\subsection{The k-NN Phenomenon}

The most striking pattern is that MiniRocket elevates k-NN from among the weakest classifiers (on Global Pooling features) to the strongest (on MiniRocket features) on \texttt{val\_outside}. Table~\ref{tab:main_results} summarizes representative numbers for this split.

\begin{table}[htbp]
\centering
\small
\caption{\textbf{Selected models on cross-species validation (\texttt{val\_outside}).} MiniRocket k-NN achieves the highest MCC (0.753), but this split-level aggregate masks species-level heterogeneity. Species-level analysis (Section~\ref{sec:neighbor_analysis}) reveals that gains concentrate in species where cassette-mediated resistance is plausible, while others show minimal improvement.}
\begin{tabular}{lcccc}
\toprule
\textbf{Method--Classifier} & \textbf{AUROC} & \textbf{AUPRC} & \textbf{F1} & \textbf{MCC} \\
\midrule
\multicolumn{5}{l}{\textit{MiniRocket features:}} \\
\quad k-NN (cos, $k=10$)     & 0.926 & 0.992 & 0.982 & \textbf{0.753} \\
\quad ExtraTrees              & 0.922 & 0.992 & 0.981 & 0.720 \\
\quad RandomForest            & 0.920 & 0.990 & 0.964 & 0.348 \\
\quad LightGBM                & 0.921 & 0.990 & 0.928 & 0.486 \\
\midrule
\multicolumn{5}{l}{\textit{Global Pooling (Baseline) features:}} \\
\quad ExtraTrees              & 0.945 & 0.988 & 0.948 & 0.263 \\
\quad RandomForest            & 0.887 & 0.978 & 0.950 & 0.323 \\
\quad LightGBM                & 0.907 & 0.983 & 0.835 & 0.418 \\
\quad Linear SVM              & 0.910 & 0.980 & 0.960 & \textbf{0.536} \\
\quad k-NN (cos, $k=10$)      & 0.515 & 0.902 & 0.901 & 0.148 \\
\bottomrule
\end{tabular}
\label{tab:main_results}
\end{table}

On Global Pooling features, k-NN dramatically underperforms linear models on \texttt{test\_outside}. After the MiniRocket transform, k-NN achieves MCC $= 0.753$—but this aggregate conceals substantial species-level variation. As shown in Section~\ref{sec:neighbor_analysis}, MiniRocket reorganizes feature space in a way that is more consistent with shared resistance modules than with phylogeny, benefiting species where cassette-mediated resistance is plausible while providing little advantage for chromosomally-mediated resistance.

\subsection{Phylogenetic Distance Analysis}

\begin{figure}[htbp]
\centering
\includegraphics[width=\textwidth]{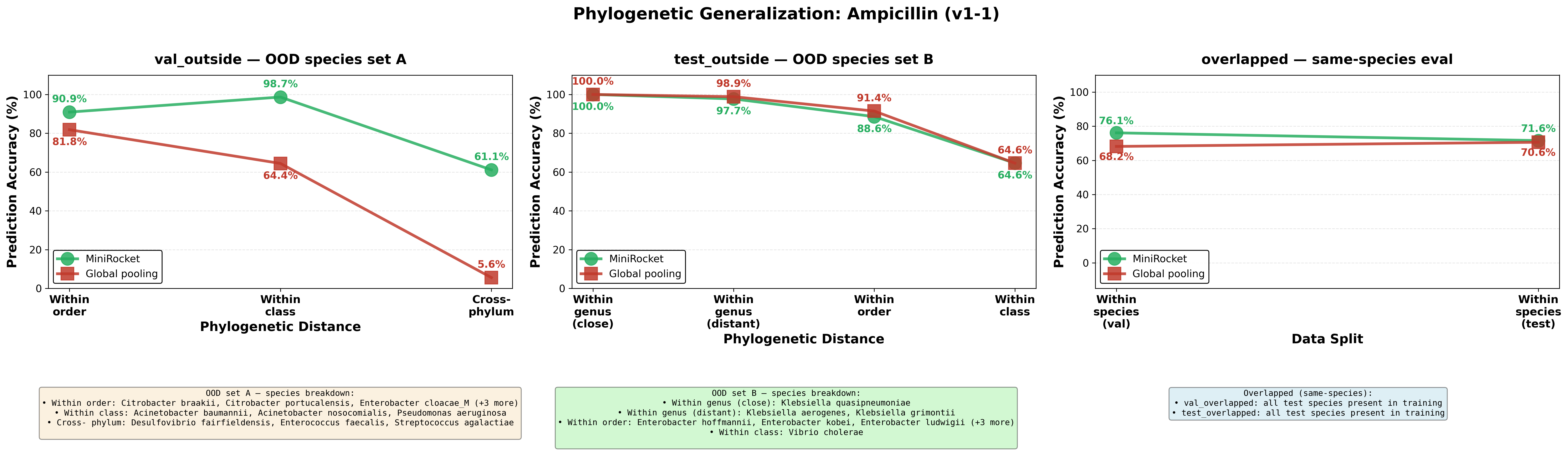}
\caption{\textbf{Phylogenetic generalization for ampicillin (replicate v1-1).} Accuracy as a function of phylogenetic distance for \texttt{val\_outside} (left), \texttt{test\_outside} (middle), and overlapped same-species splits (right). MiniRocket maintains high accuracy with increasing distance on \texttt{val\_outside}, whereas Global Pooling accuracy drops sharply, especially for cross-phylum transfer. However, on \texttt{test\_outside}, Global Pooling performs comparably or better, suggesting that phylogenetic distance alone does not determine performance—resistance mechanism is the critical factor. Both pipelines perform similarly on same-species evaluation. Replicates v1-2 and v1-3 show qualitatively similar trends (Appendix~\ref{app:phylo_replicates}). We report accuracy here for intuitive visualization; MCC shows qualitatively similar trends.}
\label{fig:phylogenetic_generalization}
\end{figure}

Figure~\ref{fig:phylogenetic_generalization} breaks down performance by phylogenetic distance for one representative replicate (v1-1). On \texttt{val\_outside}, MiniRocket maintains accuracy above 90\% within order and degrades gradually to 61.1\% for cross-phylum species, while Global Pooling drops to 5.6\% in the cross-phylum setting. The \texttt{test\_outside} split uses a different set of held-out species with different mechanistic profiles. Rather than indicating that MiniRocket fails at large phylogenetic distances, this suggests that \texttt{test\_outside} species rely more heavily on chromosomal or diffuse resistance mechanisms that do not benefit from local pattern preservation. Phylogenetic distance alone does not determine which method succeeds; resistance mechanism is likely the critical factor.

On the overlapped splits, both pipelines are stable, and their accuracies are close. This supports the view that MiniRocket is especially helpful when training and test species are phylogenetically distant, not that it universally dominates.

\section{Why k-NN Succeeds: Neighbor Analysis}
\label{sec:neighbor_analysis}
The split-dependent results in Section~\ref{sec:overall_performance}, where MiniRocket dominates \texttt{val\_outside} while Global Pooling dominates \texttt{test\_outside}, suggest that aggregating performance across entire splits obscures the true pattern. We therefore examine species-level behavior to identify which genomes benefit from MiniRocket and which do not. Neighbor analysis provides a direct window into feature space reorganization: if MiniRocket preserves local resistance-associated signals, we should observe that (1) genomes shift from phylogenetically similar neighbors to mechanistically similar neighbors, and (2) species with plausible cassette-mediated resistance show the largest accuracy gains.

\subsection{Neighbor Selection Patterns}

The success of k-NN gives a direct window into the feature space. We examine how the species composition of the 20 nearest neighbors for each test genome changes between Global Pooling and MiniRocket features, focusing here on replicate v1-1. We choose $k = 20$ for neighbor analysis to sample a broader neighborhood than would be captured with smaller values.

\begin{figure}[htbp]
\centering
\includegraphics[width=\textwidth]{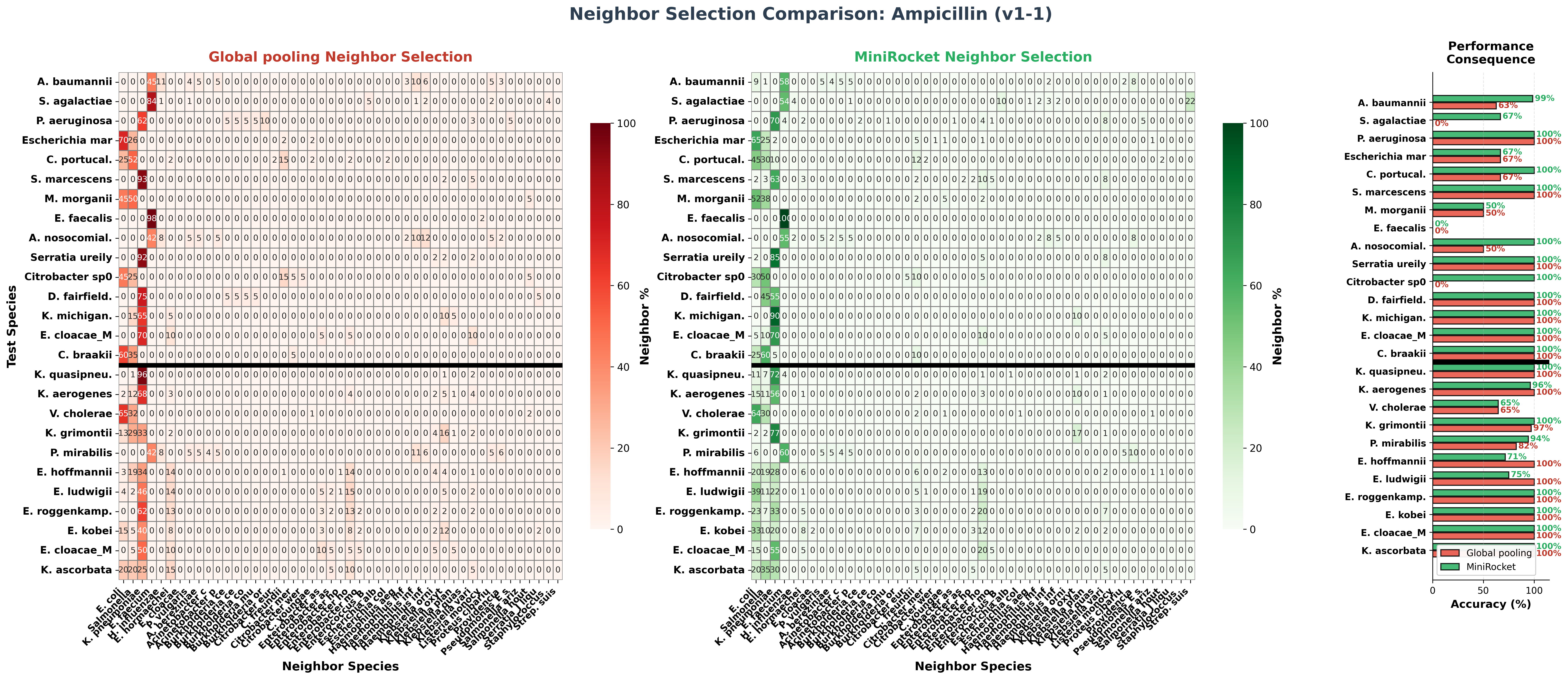}
\caption{\textbf{Neighbor selection (replicate v1-1).} Left: Global Pooling neighbor percentages by test species and neighbor species, showing strong diagonal patterns (phylogenetic bias). Middle: MiniRocket neighbor percentages, with neighbors concentrated onto a smaller set of species. Right: species-level accuracies for Global Pooling and MiniRocket.}
\label{fig:neighbor_comparison}
\end{figure}

\begin{figure}[htbp]
\centering
\includegraphics[width=\textwidth]{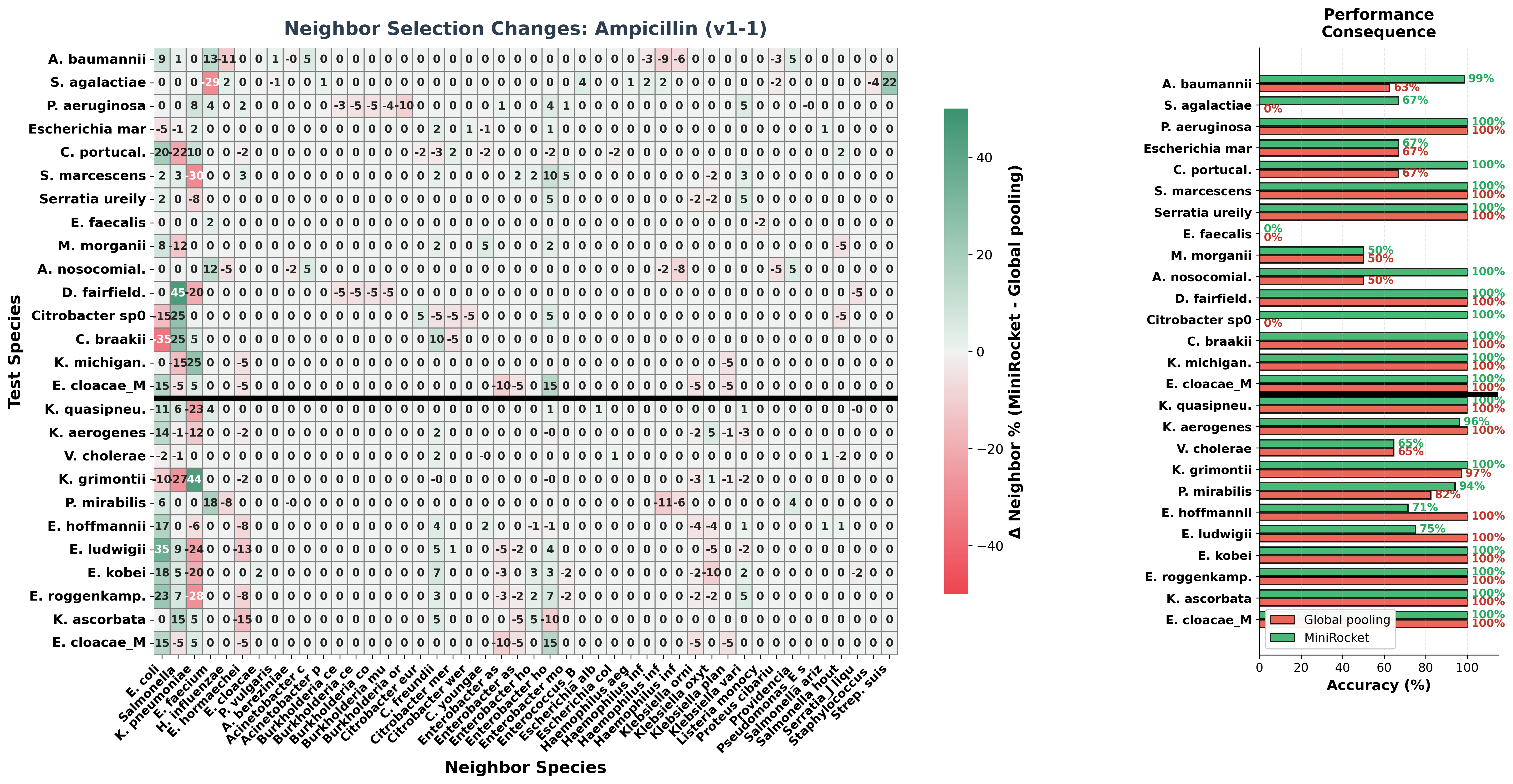}
\caption{\textbf{Neighbor selection changes (MiniRocket -- Global Pooling, replicate v1-1).} Left: change in neighbor percentages. Red cells indicate loss of neighbors; green cells indicate gain. Right: species-level accuracy for Global Pooling and MiniRocket. Species that gain neighbors from a small set of training species often show the largest accuracy improvements.}
\label{fig:neighbor_changes}
\end{figure}

Global Pooling neighbor selection (Figure~\ref{fig:neighbor_comparison}, left) is dominated by diagonal structure: test genomes mostly choose neighbors from the same or closely related species. MiniRocket (middle) reduces this diagonal and concentrates neighbors onto a smaller set of species that act as ``hubs.'' The difference heatmap in Figure~\ref{fig:neighbor_changes} makes this explicit: red blocks correspond to loss of phylogenetically close neighbors, and green blocks correspond to gains in neighbors from a few hub species.

In replicate v1-1, this reorganization is closely tied to accuracy gains. For example, \emph{Acinetobacter baumannii} accuracy increases from 63\% to 99\%, and \emph{Pseudomonas aeruginosa} from 0\% to 100\%, while their neighbors shift away from same-genus species toward a small set of training species that frequently carry ampicillin resistance. We refer to these species informally as ``AMR hubs.'' Similar patterns appear for several other species, although the magnitude of the effect varies. These neighbor analyses reveal the mechanism-dependent pattern masked by split-level aggregates: MiniRocket reorganizes feature space in a way consistent with shared resistance modules, benefiting species like \emph{A.~baumannii} and \emph{P.~aeruginosa} where cassette-mediated resistance is plausible, while providing minimal benefit to species with primarily chromosomal resistance mechanisms. The split-level inconsistencies likely reflect which mechanism profiles dominate each held-out species set, not which method is fundamentally superior.

\subsection{Interpretation and Caveats}

One natural hypothesis is that AMR hubs correspond to species enriched for transferable resistance modules, such as plasmids or genomic islands carrying $\beta$-lactamases. Under this hypothesis, MiniRocket's local pattern features make genomes that share such modules appear close in feature space, even when they are taxonomically distant. The neighbor shifts we observe are \emph{consistent} with this explanation, but they do not prove it: we have not yet performed locus-level attribution to link high-importance windows to specific cassettes or genes. A direct mechanism-composition audit (e.g., AMR gene family calls per species using CARD or ResFinder) is a natural next step; here we use neighbor reweighting and species-level performance as indirect evidence. Throughout this chapter, any references to cassette- or mechanism-based explanations should be read as hypotheses supported by neighbor and performance patterns, not as confirmed mechanistic claims.

\section{Comparison with k-mer Methods}

As discussed in Chapter~\ref{chap:kover}, k-mer-based methods such as Kover perform strongly in within-species settings but degrade under cross-species transfer, with cross-species F1 in the 0.12--0.71 range. On the same cross-species validation task considered here, both embedding pipelines outperform Kover across all cross-species scenarios: Global Pooling ExtraTrees achieves F1 $= 0.948$ on \texttt{val\_outside}, while MiniRocket k-NN achieves F1 $= 0.982$ on the same split. The key insight is not that MiniRocket universally outperforms Global Pooling, but that both contextualized embedding approaches, when matched to appropriate mechanisms, substantially outperform discrete k-mer features.

This hierarchy from k-mers, to globally pooled embeddings, to local pattern embeddings suggests that contextualized embeddings already capture functional relationships that k-mers miss, and that preserving local patterns further improves robustness to sequence variation and phylogenetic shift.

\section{The Mechanism-Mix Hypothesis}

The mechanism-mix hypothesis explains the apparent contradictions observed across evaluation splits. Within any bacterial species, resistant genomes arise from a heterogeneous mixture of mechanisms, including horizontally transferred resistance cassettes carried on plasmids or transposons, chromosomal point mutations in regulatory genes, and distributed structural changes affecting gene expression. Some mechanisms are spatially local and modular, while others are diffuse, polygenic, and tightly coupled to species-specific genomic background. As a result, AMR prediction across unseen species is effectively multi-modal in practice: the success of a given aggregation strategy depends on which resistance mechanisms dominate in the held-out species.

MiniRocket's local pattern features are well matched to cassette-mediated resistance. Plasmid-borne $\beta$-lactamases typically occupy contiguous kilobase-scale loci and are transferred horizontally across species, producing spatially localized patterns in embedding space that are conserved across taxa. By preserving positional order and summarizing local convolutional responses, MiniRocket preferentially captures these signals while reducing reliance on global genomic context. In contrast, Global Pooling aggregates statistics across the entire genome representation and is better aligned with diffuse resistance mechanisms such as efflux pump upregulation, membrane permeability changes, and chromosomal mutations distributed across multiple loci, where no single local segment dominates the phenotype.

From this perspective, the observed discrepancies between evaluation splits, with MiniRocket excelling on \texttt{val\_outside} (MCC: 0.753) and Global Pooling dominating on \texttt{test\_outside} (LightGBM MCC: 0.932), do not primarily indicate instability of the methods. Instead, they likely correspond to differences in the resistance mechanism composition of the held-out species sets. This interpretation is supported by detailed neighbor analysis across representative case studies.

\paragraph{Case 1: \emph{Proteus mirabilis} (local, variable regime).}
Resistance in \emph{P.~mirabilis} is largely driven by acquired $\beta$-lactamases, including TEM, ESBLs, and plasmid-mediated AmpC, placing it in a local but heterogeneous regime. Under Global Pooling, prediction accuracy reaches 82.4\% (14/17), but neighbor selection is dominated by a large block of \emph{Haemophilus influenzae} variants (approximately 24\%), consistent with global compositional similarity rather than resistance-mechanism similarity. MiniRocket improves accuracy to 94.1\% (16/17) and reshapes the neighborhood structure: \emph{Providencia stuartii}, a species that frequently harbors mobile $\beta$-lactamases, increases in prominence, while \emph{Haemophilus} neighbors disappear. Closely related \emph{Proteus} species also emerge, indicating a shift toward neighbors aligned by resistance mechanism rather than background genomic similarity.

\paragraph{Case 2: \emph{Streptococcus agalactiae} (chromosomal, lineage-coupled regime).}
Ampicillin resistance in \emph{S.~agalactiae} arises primarily through chromosomal variation in penicillin-binding proteins, which are highly specific at the species and genus level. Global Pooling fails entirely in this setting (0/15 correct), collapsing onto \emph{Enterococcus faecium} as a dominant hub neighbor (83.7\%), despite its intrinsic PBP5-mediated resistance being mechanistically mismatched to \emph{Streptococcus}. MiniRocket raises accuracy to 66.7\% (10/15) and introduces \emph{Streptococcus suis} as a major neighbor (22.3\%), restoring a more biologically relevant penicillin-binding protein context. Although the \emph{Enterococcus} hub remains influential, the emergence of genus-level neighbors illustrates MiniRocket's ability to recover species-relevant local patterns even when strong global signals are present.

\paragraph{Case 3: \emph{Acinetobacter baumannii} (local, acquired regime).}
Resistance in \emph{A.~baumannii} is driven by a combination of intrinsic OXA-51 and acquired OXA-23 or OXA-24 $\beta$-lactamases. Global Pooling yields moderate accuracy (62.6\%) but produces neighborhoods dominated by \emph{Haemophilus influenzae} variants (approximately 27\%), which share neither resistance mechanism nor regulatory context with \emph{Acinetobacter}. MiniRocket dramatically improves accuracy to 98.6\% (277/281) and reorganizes neighborhoods around species with known acquired $\beta$-lactamases, including \emph{Escherichia coli}, \emph{Providencia stuartii}, and closely related \emph{Acinetobacter} species carrying OXA-type enzymes. The disappearance of \emph{Haemophilus} neighbors highlights MiniRocket's tendency to reduce reliance on global compositional similarity in favor of mechanistic alignment.

\paragraph{Case 4: \emph{Enterobacter hoffmannii} (global, conserved regime).}
In contrast, \emph{E.~hoffmannii} belongs to an Enterobacter clade in which ampicillin phenotypes are often dominated by conserved chromosomal background (e.g., inducible \textit{ampC}). Global Pooling achieves perfect accuracy (7/7), with neighbors drawn almost exclusively from Enterobacterales, including \emph{Klebsiella pneumoniae}, \emph{Salmonella enterica}, and other \emph{Enterobacter} species sharing a common chromosomal background. MiniRocket reduces accuracy to 71.4\% and introduces off-clade neighbors such as \emph{Escherichia coli}, diluting the conserved global signal. In this setting, preserving locality offers little advantage and can be detrimental by fragmenting an otherwise informative global background.

Taken together, these case studies demonstrate that no single aggregation strategy is universally superior. MiniRocket tends to excel when resistance is plausibly driven by local, horizontally transferable elements, whereas Global Pooling is better aligned with conserved chromosomal mechanisms that manifest as broad genomic patterns. Performance differences across evaluation splits therefore likely reflect the mixture of resistance mechanisms present in the held-out species rather than an intrinsic advantage of one method over the other. This mechanism-dependent behavior underscores the importance of interpreting AMR benchmarks through a biological lens rather than relying solely on aggregate performance metrics.

\section{Limitations and Future Work}

Several limitations qualify the conclusions above.

First, our analysis focuses on a single antibiotic (ampicillin). Other antibiotics with different mechanistic profiles (for example, fluoroquinolones dominated by chromosomal mutations) may benefit less from locality preservation, and some may require different aggregation schemes entirely.

Second, although we design the splits to separate species, the \texttt{test\_outside} and \texttt{val\_outside} partitions may still differ in intrinsic difficulty and mechanism composition. We mitigate this by emphasizing within-split comparisons and by using multiple metrics, but observational genomic data rarely permit perfect experimental control.

Third, our mechanistic interpretations rely on indirect evidence from neighbors and performance. Future work should connect MiniRocket features to specific genomic loci via attribution maps intersected with AMR gene annotations (e.g., CARD or ResFinder calls), ablation or masking experiments, and, ideally, experimental validation.

\section{Practical Implications}

\subsection{Design and Computational Trade-offs}

The Global Pooling and MiniRocket pipelines represent a trade-off between simplicity and cross-species robustness.

Global Pooling uses 246 features per genome and works well on same-species evaluation. It is simple to implement, fast to train, and may be adequate when deployment species are similar to the training set or when computational resources are limited.

MiniRocket produces about 12{,}000 features per genome, increasing memory footprint and classifier training time, but requires no additional trainable parameters in the feature extractor itself. Our experiments show that this extra cost pays off specifically for species where cassette-mediated resistance is plausible. Practitioners should choose aggregation strategy based on expected resistance mechanisms: MiniRocket for pathogens known to carry plasmid-borne resistance (e.g., carbapenem-resistant Enterobacteriaceae), Global Pooling for species with chromosomal resistance (e.g., fluoroquinolone resistance via \textit{gyrA} mutations), and hybrids when mechanism profiles are unknown.

\subsection{Using Classifiers as Diagnostics}

Classifier behavior itself can serve as a diagnostic. When k-NN performs poorly and complex models such as gradient-boosted trees dominate, the feature space likely mixes phenotype and confounders such as species composition. When a simple k-NN with cosine distance outperforms sophisticated ensembles, as with MiniRocket on \texttt{val\_outside}, it suggests that the representation has meaningful local structure and that neighbor inspection can be used for explanation and data auditing.

\section{Conclusion}

Treating Evo embeddings as ordered signal streams and summarizing them with MiniRocket fundamentally changes how ampicillin resistance is represented, but the benefit is mechanism-dependent, not universal. Initial split-level results appeared contradictory: MiniRocket k-NN achieved MCC $= 0.753$ on \texttt{val\_outside} vs.\ Global Pooling's 0.148, yet Global Pooling dominated \texttt{test\_outside} (LightGBM: 0.932 vs.\ 0.798). This inconsistency motivated species-level analysis, which revealed the underlying pattern: MiniRocket excels for species where cassette-mediated resistance is plausible (e.g., \emph{A.~baumannii}, \emph{P.~aeruginosa}), while providing minimal benefit, or even degraded performance, for species with chromosomal or diffuse mechanisms.

Neighbor analyses show that MiniRocket reduces phylogenetic bias and redirects neighbors toward a smaller set of species that act as AMR hubs, and phylogenetic distance plots indicate that MiniRocket degrades more gracefully with evolutionary distance on \texttt{val\_outside}. However, performance on same-species splits remains comparable between methods, and Global Pooling outperforms MiniRocket on \texttt{test\_outside}, indicating that resistance mechanism profile, rather than phylogenetic distance or split identity, likely determines which aggregation strategy succeeds.

More broadly, this case study illustrates that \emph{aggregation strategy should be matched to biological mechanism}. Split-level aggregates can be misleading: they reflect which mechanisms dominate each held-out species set, not which method is fundamentally superior. Species-level analysis reveals the true pattern: local pattern preservation benefits species with plausible cassette-dominated resistance, while global pooling handles diffuse or chromosomal mechanisms more robustly. This mechanism-dependent performance explains both why MiniRocket dramatically improved some species' accuracy (those with likely cassette-based resistance) and why it provided no benefit for others.

Overall, our results suggest a mechanistic design principle for genomic foundation model aggregation: \emph{match aggregation strategy to resistance mechanism}. For cassette-mediated resistance (plasmids, transposons, genomic islands), treat embeddings as ordered signals and apply local pattern extractors like MiniRocket, which reorganize feature space to cluster genomes by shared modules rather than phylogeny. For chromosomal or diffuse resistance (point mutations, efflux pumps, membrane changes), use global pooling with tree-based classifiers, which better capture genome-wide compositional signals. When mechanism profiles are unknown or mixed, ensemble both approaches. The key insight is that split-level performance aggregates can be artifacts of species composition—species-level analysis is required to understand when and why each method succeeds.

%% file: Conclusion.tex
\chapter{Conclusion}
\label{ch:conclusion}

This thesis addressed cross-species antimicrobial resistance (AMR) prediction using genomic foundation models. We developed a diagnostic-driven framework for embedding extraction and investigated how aggregation strategy interacts with resistance mechanism to determine generalization performance.

\section{Contributions}

This work makes four contributions. First, we designed a species holdout evaluation protocol that enforces zero phylogenetic overlap between training and test sets, revealing generalization failures masked by standard random-split evaluation. Second, we developed a diagnostic framework for layer selection in genomic foundation models, identifying Layer 10 as the deepest stable extraction point in Evo-1-8k-base under native bfloat16 inference. Third, we introduced MiniRocket-based aggregation that treats genomic embeddings as ordered signals, preserving cassette-scale patterns that Global Pooling dilutes. Fourth, we provided empirical evidence for a mechanism-mix hypothesis: MiniRocket excels when cassette-mediated resistance predominates, while Global Pooling remains competitive for chromosomal or diffuse mechanisms.

\section{Key Takeaways}

Across aggregation methods and evaluation splits, Evo-based embeddings consistently outperform the k-mer baseline (Kover; cross-species F1 often below 0.70), confirming that foundation model representations enable stronger cross-species generalization than discrete sequence features. Both Global Pooling and MiniRocket achieve high accuracy, with neither method universally dominating. Performance depends on resistance mechanism composition of held-out species: local pattern preservation benefits species with cassette-dominated resistance, while Global Pooling handles diffuse or chromosomal mechanisms more robustly. After MiniRocket transformation, simple k-NN classifiers become top performers on certain splits, indicating that geometric reorganization of feature space, not classifier complexity, drives cross-species generalization.

\section{Limitations}

Several limitations qualify these conclusions. Our detailed analyses focused on ampicillin resistance; other antibiotics with different mechanistic profiles may show different patterns. Both representations exhibit hubness, where certain species (e.g., \textit{Enterococcus faecium}) appear disproportionately as neighbors, potentially introducing systematic biases. The observed alignment between neighbor geometry and performance is correlational, not causal. Our mechanistic interpretations rely on indirect evidence from neighbor analysis rather than direct locus-level attribution. All analyses used retrospectively collected data; prospective clinical validation is required before deployment. Finally, we evaluated only Evo-1-8k-base; other genomic foundation models may behave differently.

\section{Future Directions}

Intersecting high-importance MiniRocket features with AMR gene annotations (e.g., CARD, ResFinder) could validate whether the method genuinely captures resistance cassettes. Systematic evaluation across antibiotics with diverse resistance mechanisms would further test the mechanism-mix hypothesis. Investigating hubness-reduction techniques could improve k-NN reliability. Clinical deployment on prospectively collected isolates would assess real-world utility.

\section{Closing Remarks}

Cross-species AMR prediction is constrained by the heterogeneous biology of resistance. No single approach will universally succeed because resistance arises from diverse mechanisms with different transferability properties. This thesis demonstrates that matching aggregation strategy to biological mechanism enables more effective foundation model deployment. The broader lesson is that effective use of large pretrained models in biology requires understanding both their computational properties and the biological structure of the prediction task.

%% file: appendix_phylo.tex
\appendix

\chapter{Phylogenetic Generalization Across Replicates}
\label{app:phylo_replicates}

This appendix presents phylogenetic generalization results from replicates v1-2 and v1-3 to complement the v1-1 results presented in Chapter 5. Consistent patterns across all three independent replicates confirm the robustness of the observed trends.

\section{Replicate v1-2}

\begin{figure}[h]
\centering
\includegraphics[width=\textwidth]{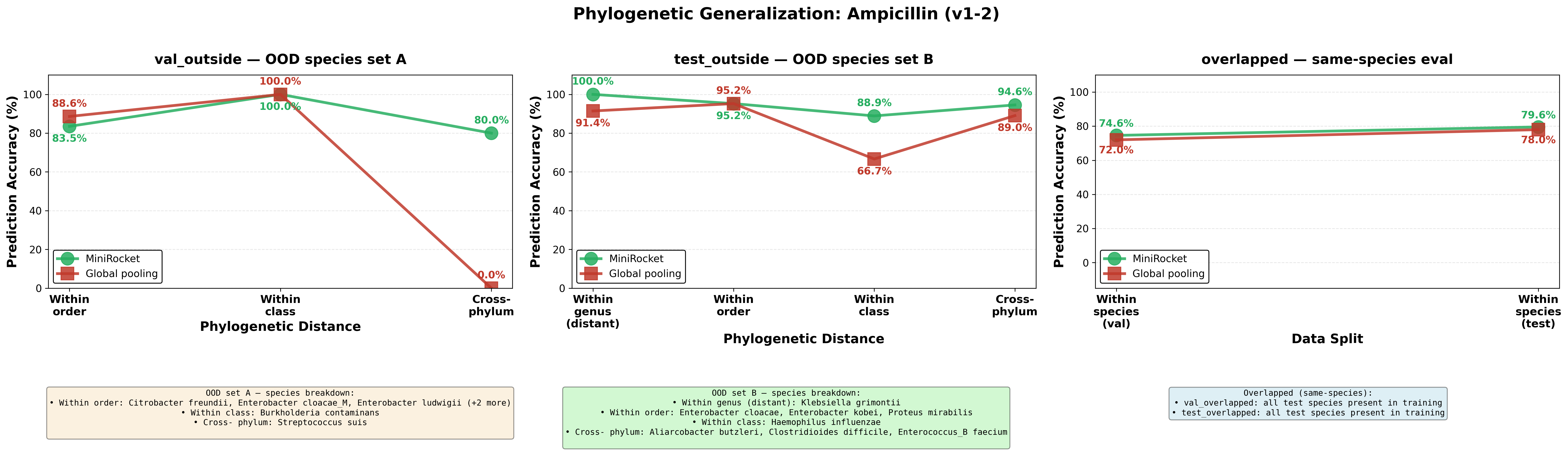}
\caption{\textbf{Phylogenetic generalization across evolutionary distance (replicate v1-2).} MiniRocket (green) maintains strong performance across phylogenetic distances on \texttt{val\_outside}, while Global Pooling (red) shows catastrophic cross-phylum collapse. Both methods perform comparably on \texttt{test\_outside} for this replicate.}
\label{fig:phylo_v1_2}
\end{figure}

Figure \ref{fig:phylo_v1_2} shows phylogenetic generalization patterns for replicate v1-2. On \texttt{val\_outside}, both methods start with comparable performance at close phylogenetic distances, but MiniRocket maintains substantially better accuracy at cross-phylum distance, while Global Pooling shows complete performance collapse. The \texttt{test\_outside} split shows both methods achieving generally strong performance across phylogenetic distances, indicating that this replicate's held-out test species were less challenging. On overlapped splits, both methods achieve comparable accuracy.

\section{Replicate v1-3}

\begin{figure}[h]
\centering
\includegraphics[width=\textwidth]{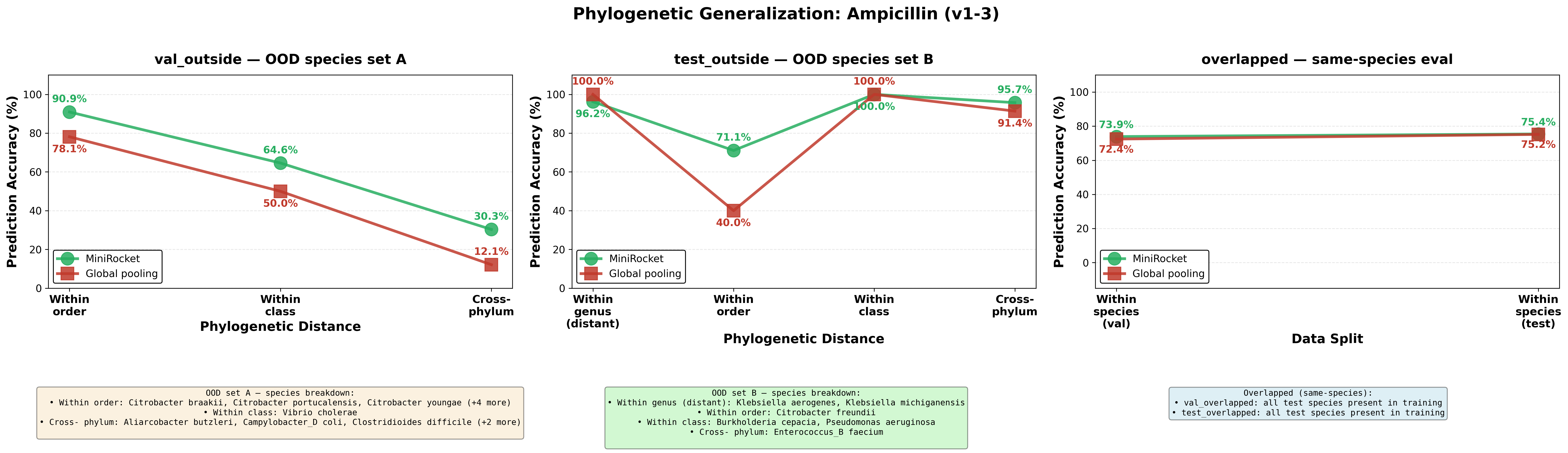}
\caption{\textbf{Phylogenetic generalization across evolutionary distance (replicate v1-3).} MiniRocket (green) shows more pronounced degradation on \texttt{val\_outside} compared to v1-1 and v1-2, but maintains advantage over Global Pooling (red). Test set performance remains strong across phylogenetic distances.}
\label{fig:phylo_v1_3}
\end{figure}

Figure \ref{fig:phylo_v1_3} shows phylogenetic generalization patterns for replicate v1-3. On \texttt{val\_outside}, MiniRocket shows more pronounced degradation with increasing phylogenetic distance compared to v1-1 and v1-2, but maintains a consistent advantage over Global Pooling at all distances. The cross-phylum gap between methods remains substantial. On \texttt{test\_outside}, MiniRocket maintains strong performance across all phylogenetic distances, while Global Pooling shows more variable results. Overlapped splits show comparable performance between methods.

%% file: references.bib
@techreport{WHO2019,
  author = {{World Health Organization}},
  title = {No Time to Wait: Securing the Future from Drug-Resistant Infections},
  institution = {Interagency Coordination Group on Antimicrobial Resistance},
  year = {2019},
  url = {https://www.who.int/antimicrobial-resistance/interagency-coordination-group/final-report/en/}
}

@techreport{ONeill2016,
  author = {O'Neill, Jim},
  title = {Tackling Drug-Resistant Infections Globally: Final Report and Recommendations},
  institution = {The Review on Antimicrobial Resistance},
  year = {2016},
  url = {https://amr-review.org/sites/default/files/160525_Final%20paper_with%20cover.pdf}
}

@manual{CLSI2020,
  author = {{Clinical and Laboratory Standards Institute}},
  title = {Performance Standards for Antimicrobial Susceptibility Testing},
  organization = {CLSI},
  edition = {30th},
  year = {2020},
  address = {Wayne, PA},
  note = {CLSI supplement M100}
}

@article{Murray2022,
  author = {Murray, Christopher J. L. and Ikuta, Kevin Shunji and Sharara, Fablina and Swetschinski, Lucien and {Robles Aguilar}, Gisela and Gray, Authia and Han, Chieh and Bisignano, Catherine and Rao, Puja and Wool, Eve and others},
  title = {Global burden of bacterial antimicrobial resistance in 2019: a systematic analysis},
  journal = {The Lancet},
  volume = {399},
  number = {10325},
  pages = {629--655},
  year = {2022},
  doi = {10.1016/S0140-6736(21)02724-0}
}

@article{Nishida2012,
  author = {Nishida, Hiromi},
  title = {Comparative analyses of base compositions, DNA sizes, and dinucleotide frequency profiles in archaeal and bacterial chromosomes and plasmids},
  journal = {International Journal of Evolutionary Biology},
  volume = {2012},
  pages = {342482},
  year = {2012},
  doi = {10.1155/2012/342482}
}

@inproceedings{nguyen2023hyenadna,
  author    = {Nguyen, Eric and Poli, Michael and Faizi, Marjan and Thomas, Armin and Birch, Michael and Wornow, Michael and Patel, Aman and Rabideau, Clayton and Massaroli, Stefano and Bengio, Yoshua and Ermon, Stefano and Baccus, Stephen A. and R{\'e}, Christopher},
  title     = {{HyenaDNA}: Long-Range Genomic Sequence Modeling at Single Nucleotide Resolution},
  booktitle = {Advances in Neural Information Processing Systems 36 (NeurIPS)},
  year      = {2023},
  url       = {https://arxiv.org/abs/2306.15794}
}

@article{Drouin2016,
  author = {Drouin, Alexandre and Giguère, Sébastien and Déraspe, Maxime and Marchand, Mario and Tyers, Michael and Loo, Vivian G. and Bourgault, Anne-Marie and Laviolette, François and Corbeil, Jacques},
  title = {Predictive computational phenotyping and biomarker discovery using reference-free genome comparisons},
  journal = {BMC Genomics},
  volume = {17},
  number = {1},
  pages = {754},
  year = {2016},
  doi = {10.1186/s12864-016-2889-6}
}

@article{Drouin2019,
  author = {Drouin, Alexandre and Letarte, Gaël and Raymond, Frédéric and Marchand, Mario and Corbeil, Jacques and Laviolette, François},
  title = {Interpretable genotype-to-phenotype classifiers with performance guarantees},
  journal = {Scientific Reports},
  volume = {9},
  number = {1},
  pages = {4071},
  year = {2019},
  doi = {10.1038/s41598-019-40561-2}
}

@article{Hu2024,
  author = {Hu, Kaixin and Meyer, Fernando and Deng, Zhi-Luo and Asgari, Ehsaneddin and Kuo, Tzu-Hao and Münch, Philipp C. and McHardy, Alice C.},
  title = {Assessing computational predictions of antimicrobial resistance phenotypes from microbial genomes},
  journal = {Briefings in Bioinformatics},
  volume = {25},
  number = {3},
  pages = {bbae206},
  year = {2024},
  doi = {10.1093/bib/bbae206}
}

@article{Nguyen2024,
  author = {Nguyen, Eric and Poli, Michael and Durrant, Matthew G. and Kang, Brian and Katrekar, Dhruva and Li, David B. and Bartie, Liam J. and Thomas, Armin W. and King, Samuel H. and Brixi, Garyk and Sullivan, Jeremy and Ng, Madelena Y. and Lewis, Ashley and Lou, Aaron and Ermon, Stefano and Baccus, Stephen A. and Hernandez-Boussard, Tina and Ré, Christopher and Hsu, Patrick D. and Hie, Brian L.},
  title = {Sequence Modeling and Design from Molecular to Genome Scale with {Evo}},
  journal = {Science},
  volume = {386},
  number = {6723},
  pages = {eado9336},
  year = {2024},
  doi = {10.1126/science.ado9336}
}

@article{Ji2021,
  author = {Ji, Yanrong and Zhou, Zhihan and Liu, Han and Davuluri, Ramana V.},
  title = {{DNABERT}: pre-trained Bidirectional Encoder Representations from Transformers model for {DNA}-language in genome},
  journal = {Bioinformatics},
  volume = {37},
  number = {15},
  pages = {2112--2120},
  year = {2021},
  doi = {10.1093/bioinformatics/btab083}
}

@article{Zhou2023,
  author = {Zhou, Zhihan and Ji, Yanrong and Li, Weijian and Dutta, Pratik and Davuluri, Ramana and Liu, Han},
  title = {{DNABERT-2}: Efficient Foundation Model and Benchmark For Multi-Species Genome},
  journal = {arXiv preprint arXiv:2306.15006},
  year = {2023},
  url = {https://arxiv.org/abs/2306.15006}
}

@article{DallaTorre2023,
  author = {Dalla-Torre, Hugo and Gonzalez, Liam and Mendoza Revilla, Javier and Lopez Carranza, Nicolas and Henryk Grywaczewski, Adam and Oteri, Francesco and Dallago, Christian and Trop, Evan and Sirelkhatim, Hassan and Richard, Guillaume and others},
  title = {The Nucleotide Transformer: Building and Evaluating Robust Foundation Models for Human Genomics},
  journal = {bioRxiv},
  year = {2023},
  doi = {10.1101/2023.01.11.523679}
}

@article{Dempster2020,
  author = {Dempster, Angus and Petitjean, François and Webb, Geoffrey I.},
  title = {{ROCKET}: Exceptionally fast and accurate time series classification using random convolutional kernels},
  journal = {Data Mining and Knowledge Discovery},
  volume = {34},
  number = {5},
  pages = {1454--1495},
  year = {2020},
  doi = {10.1007/s10618-020-00701-z}
}

@inproceedings{Dempster2021,
  author = {Dempster, Angus and Schmidt, Daniel F. and Webb, Geoffrey I.},
  title = {{MINIROCKET}: A Very Fast (Almost) Deterministic Transform for Time Series Classification},
  booktitle = {Proceedings of the 27th ACM SIGKDD Conference on Knowledge Discovery \& Data Mining},
  pages = {248--257},
  year = {2021},
  doi = {10.1145/3447548.3467231}
}

@article{Shimodaira2000,
  author = {Shimodaira, Hidetoshi},
  title = {Improving predictive inference under covariate shift by weighting the log-likelihood function},
  journal = {Journal of Statistical Planning and Inference},
  volume = {90},
  number = {2},
  pages = {227--244},
  year = {2000},
  doi = {10.1016/S0378-3758(00)00115-4}
}

@inproceedings{tenney2019bert,
  author    = {Tenney, Ian and Das, Dipanjan and Pavlick, Ellie},
  title     = {{BERT} Rediscovers the Classical {NLP} Pipeline},
  booktitle = {Proceedings of the 57th Annual Meeting of the Association for Computational Linguistics (ACL)},
  pages     = {4593--4601},
  year      = {2019},
  address   = {Florence, Italy},
  publisher = {Association for Computational Linguistics},
  doi       = {10.18653/v1/P19-1452}
}

@article{rogers2020primer,
  author  = {Rogers, Anna and Kovaleva, Olga and Rumshisky, Anna},
  title   = {A Primer in {BERT}ology: What We Know About How {BERT} Works},
  journal = {Transactions of the Association for Computational Linguistics},
  volume  = {8},
  pages   = {842--866},
  year    = {2020},
  doi     = {10.1162/tacl_a_00349}
}

@inproceedings{yosinski2014transfer,
  author    = {Yosinski, Jason and Clune, Jeff and Bengio, Yoshua and Lipson, Hod},
  title     = {How Transferable Are Features in Deep Neural Networks?},
  booktitle = {Advances in Neural Information Processing Systems 27 (NeurIPS)},
  pages     = {3320--3328},
  year      = {2014},
  url       = {https://papers.nips.cc/paper_files/paper/2014/hash/532a2f85b6977104bc93f8580abbb330-Abstract.html}
}

@inproceedings{raghu2019transfusion,
  author    = {Raghu, Maithra and Zhang, Chiyuan and Kleinberg, Jon and Bengio, Samy},
  title     = {Transfusion: Understanding Transfer Learning for Medical Imaging},
  booktitle = {Advances in Neural Information Processing Systems (NeurIPS)},
  volume    = {32},
  year      = {2019}
}

@inproceedings{kornblith2019similarity,
  author    = {Kornblith, Simon and Norouzi, Mohammad and Lee, Honglak and Hinton, Geoffrey},
  title     = {Similarity of Neural Network Representations Revisited},
  booktitle = {Proceedings of the 36th International Conference on Machine Learning (ICML)},
  series    = {Proceedings of Machine Learning Research},
  volume    = {97},
  pages     = {3519--3529},
  year      = {2019},
  publisher = {PMLR}
}

@inproceedings{ethayarajh2019contextual,
  author    = {Ethayarajh, Kawin},
  title     = {How Contextual Are Contextualized Word Representations? Comparing the Geometry of {BERT}, {ELMo}, and {GPT-2} Embeddings},
  booktitle = {Proceedings of EMNLP-IJCNLP},
  pages     = {55--65},
  year      = {2019},
  address   = {Hong Kong, China},
  publisher = {Association for Computational Linguistics},
  doi       = {10.18653/v1/D19-1006}
}

@article{micikevicius2017mixed,
  author  = {Micikevicius, Paulius and Narang, Sharan and Alben, Jonah and Diamos, Gregory and Elsen, Erich and Garcia, David and Ginsburg, Boris and Houston, Michael and Kuchaiev, Oleksii and Venkatesh, Ganesh and Wu, Hao},
  title   = {Mixed Precision Training},
  journal = {arXiv preprint arXiv:1710.03740},
  year    = {2017},
  url     = {https://arxiv.org/abs/1710.03740}
}

@article{queipo2025sinks,
  author  = {Queipo-de-Llano, Enrique and Arroyo, {\'A}lvaro and Barbero, Federico and Dong, Xiaowen and Bronstein, Michael and LeCun, Yann and Shwartz-Ziv, Ravid},
  title   = {Attention Sinks and Compression Valleys in {LLMs} Are Two Sides of the Same Coin},
  journal = {arXiv preprint arXiv:2510.06477},
  year    = {2025},
  url     = {https://arxiv.org/abs/2510.06477}
}
